\documentclass[usenatbib]{mn2e}

\usepackage{epsfig}
\usepackage{amssymb}
\usepackage{graphics}
\usepackage{afterpage}
\usepackage{hyperref}
\usepackage{xspace}
\usepackage{times}
\newcommand{\msun}{{\rm{M}_\odot}}
\newcommand{\rhocr}{\rho_{\rm{cr}}}

\newcommand{\lcdm}{$\Lambda$CDM\xspace}
\newcommand{\nbody}{$N$-body\xspace}
\newcommand{\eagle}{\textsc{eagle}\xspace}

\newcommand{\rms}{\textsc{rms}\xspace}
\newcommand{\dmonly}{\textsc{dmo}\xspace}
\newcommand{\gadget}{\textsc{gadget}\xspace}
\newcommand{\subfind}{\textsc{subfind}\xspace}
\newcommand{\anarchy}{\textsc{anarchy}\xspace}
\title[Baryon effects on the internal structure of \lcdm halos]{Baryon
  effects on the internal structure of \lcdm halos in the EAGLE simulations}

\author[M. Schaller et al.]  {Matthieu Schaller$^1$\thanks{E-mail: matthieu.schaller@durham.ac.uk}, 
			      Carlos S. Frenk$^1$,
			      Richard G. Bower$^1$,
			      Tom Theuns$^{1,2}$, \newauthor
			      Adrian Jenkins$^1$, 
			      Joop Schaye$^3$,
			      Robert A. Crain$^{3,4}$,
			      Michelle Furlong$^1$,\newauthor
			      Claudio Dalla Vecchia$^{5,6}$ and
			      I. G. McCarthy$^4$\\			     
$^1$Institute for Computational Cosmology, Durham University, South Road, Durham, UK, DH1 3LE\\
$^2$Department of Physics, University of Antwerp, Campus Groenenborger, Groenenborgerlaan 171, B-2020 Antwerp, Belgium\\
$^3$Leiden Observatory, Leiden University, P.O. Box 9513, 2300 RA Leiden, The Netherlands\\
$^4$Astrophysics Research Institute, Liverpool John Moores University, 146 Brownlow Hill, Liverpool L3 5RF, UK\\
$^5$Instituto de Astrof\'isica de Canarias, C/ V\'ia L\'actea s/n, 38205 La Laguna, Tenerife, Spain\\
$^6$Departamento de Astrof\'isica, Universidad de La Laguna, Av. del Astrof\'isico Franciso S\'anchez s/n, 38206 
La Laguna, Tenerife, Spain\\
}

\begin{document}

\date{\today}

\pagerange{\pageref{firstpage}--\pageref{lastpage}} \pubyear{2014}

\maketitle

\label{firstpage}

\begin{abstract}
  We investigate the internal structure and density profiles of halos
  of mass $10^{10}-10^{14}~\msun$ in the Evolution and Assembly of
  Galaxies and their Environment (\eagle) simulations. These follow
  the formation of galaxies in a \lcdm Universe and include a
  treatment of the baryon physics thought to be relevant. The \eagle
  simulations reproduce the observed present-day galaxy stellar mass function,
  as well as many other properties of the galaxy population as a
  function of time. We find significant differences between the masses
  of halos in the \eagle simulations and in simulations that follow
  only the dark matter component. Nevertheless, halos are well described by the
  Navarro-Frenk-White (NFW) density profile at radii larger than $\sim
  5\%$ of the virial radius but, closer to the centre, the presence of
  stars can produce cuspier profiles. Central enhancements in the
  total mass profile are most important in halos of mass $
  10^{12}-10^{13}\msun$, where the stellar fraction peaks. Over the radial
  range where they are well resolved, the resulting galaxy rotation
  curves are in very good agreement with observational data for
  galaxies with stellar mass $M_*<5\times10^{10}\msun$.  We present an
  empirical fitting function that describes the total mass profiles
  and show that its parameters are strongly correlated with halo mass.
\end{abstract}

\begin{keywords}
cosmology: theory, dark matter, large-scale structure of Universe
\end{keywords}

%################################################################################################
\section{Introduction}
\label{sec:introduction}

The development of efficient computational techniques and the growing
availablity of computing power over the past three decades have made it
possible to simulate the evolution of representative cosmological
volumes at high enough resolution to follow the formation of cosmic
structures over many orders of magnitude in mass.  

One of the best established and most robust results from this
programme is the characterization of the density structure of dark
matter (DM) halos in equilibrium whose spherically averaged density
profile, $\rho(r)$, is nearly universal in shape and obeys simple
scaling relations \citep{Navarro1996a,Navarro1997}. The functional
form of this ``NFW'' radial profile is independent of mass, formation
redshift, and cosmological parameters and has the form:

\begin{equation}
 \frac{\rho(r)}{\rhocr} = \frac{\delta_{\rm{c}}}{\left(r/r_{\rm{s}}\right)\left(1+r/r_{\rm{s}}\right)^2},
  \label{eq:nfw}
\end{equation}
where $\rhocr$ is the critical density of the Universe,
$\delta_{\rm{c}}$ a characteristic density and $r_{\rm{s}}$ 
a characteristic radius. \cite{Navarro1997} showed that these two scale
parameters are strongly correlated and that the characteristic density
is proportional to the density of the universe at the time when the
halo was assembled. This proportionality constant or, equivalently,
the proportionality constant between halo mass and concentration has
been studied by many authors \citep[e.g.][]{AvilaReese1999, Jing2000,
  Bullock2001, Eke2001, Zhao2003,Neto2007, Duffy2008, Gao2008, Navarro2010,
  Ludlow2014, Dutton2014}.  The validity of the model is well
established and a physical understanding of the universality of the
profile is beginning to emerge \citep{Ludlow2013,Correa2014,Correa2015}.

The nearly scale-free behaviour induced by gravity applies only to
halos made entirely of DM. In practice, halos of mass above $\sim 10^9
~\msun$ participate in the process of galaxy formation. The cooling
and dissipation of gas in these halos introduces a characteristic
scale that breaks self-similarity \citep{White1978, WhiteFrenk1991}
and the subsequent formation of stars can deepen the potential well and
modify the structure of the halo in this region.

One of the early models of the effects of baryon collapse on the
structure of a halo, making use of adiabatic invariants, concluded
that halos would become denser in their centres
\citep{Blumenthal1986}. These simple models, however, were later
shown not to match hydrodynamic simulations and led to a more general
framework for calculating adiabatic contraction based on the average
radial distribution of particles \citep{Gnedin2004,Gustafsson2006}.
The parameters of this model, however, have been shown to depend on
halo mass, redshift and on the details of the hydrodynamic simulation,
making analytical descriptions of adiabatic contraction complex and
uncertain \citep{Duffy2010}.

Baryons, however, can also produce the opposite effect and induce
expansion rather than contraction of the halo. Using idealized
hydrodynamic simulations, \cite{Navarro1996b} showed that the rapid
expulsion of gas that had previously cooled to very high density near
the centre of a halo could generate a central core. Subsequent work
using cosmological simulations has confirmed and extended this result
\citep[e.g.][]{Read2005, Dehnen2005,
  Mashchenko2006,Governato2010,PontzenGovernato2012,Teyssier2013,Martizzi2013}.

The structure of the inner halo is often used as a test of the 
\lcdm paradigm \citep[e.g.][]{Sand2002,Gilmore2007}.  Such tests,
however, tend to compare observations of halos which have galaxies
within them with results from simulations of pure dark matter halos
\citep{Newman2013a}. For the tests to be meaningful, accurate and
reliable calculations of how baryons affect the structure of the halos
are essential. Such calculations are also of major importance for
efforts to detect DM experimentally, either directly in the laboratory,
or indirectly through the products of particle decay or annihilation.

Simulating the evolution of the visible components of the universe is
a much more complex task than simulating the evolution of the DM
because baryons undergo a variety of astrophysical processes many of
which are relatively poorly understood. The resolution that is
attainable even with the largest computers today is insufficient for
an {\em ab initio} calculation of most of these processes which, as a
result, need to be treated through parametrized ``subgrid'' models
added to the coupled hydrodynamical and gravitational evolution
equations. These models describe the effects of radiative cooling,
star formation, feedback from energy liberated during the evolution of
stars and supermassive black holes growing at the centres of
galaxies. Simulations that include some or all of these processes have
shown that significant changes can be induced in the total masses of
halos \citep{Sawala2013, Sawala2014, Cusworth2013, Velliscig2014,
  Vogelsberger2014} and in their inner structure
\citep[e.g.][]{Gnedin2004, Pedrosa2009, Duffy2010,
  PontzenGovernato2012, Brook2012, DiCintio2014}.

In this paper we investigate how baryon physics modifies the structure
of DM halos in the Evolution and Assembly of Galaxies and their
Environment (\eagle) cosmological hydrodynamical simulations \citep{Schaye2014}. An
important advantage of these simulations is that they give a good
match to the stellar mass function and and to the distribution of
galaxy sizes over a large range of stellar masses ($
(10^{8} -10^{11.5})~\msun $). Furthermore, the
relatively large volume of the reference \eagle simulation provides a
large statistical sample to derive the halo mass function in the mass
range $(10^9-10^{14})~\msun$ and to investigate the radial density
profiles of halos more massive than $10^{11}\msun$.

This paper is organised as follows. In Section~\ref{sec:simulation} we
introduce the simulations and describe the selection of halos. In
Section~\ref{sec:HaloCounts} we focus on the change in the mass of 
halos induced by baryon processes and the effect this has on the
halo mass function. In Section~ \ref{sec:HaloProfile} we analyse the
radial density profile of the halos and decompose them according to
their different constituents. We fit the total matter profile with a
universal formula that accounts for deviations from the NFW
profile and show that the best fit parameters of these fits correlate with
the mass of the halo.  Our main results are summarized in
Section~\ref{sec:conclusion}. All our results are restricted to redshift $z=0$ and
all quantities are given in physical units (without factors of $h$).

%################################################################################################
\section{The simulations}
\label{sec:simulation}

The simulations analysed in this paper were run as part of a Virgo
Consortium project called the Evolution and Assembly of Galaxies and
their Environment (\eagle; \citealt{Schaye2014}).  The \eagle project
consists of simulations of \lcdm cosmological volumes with sufficient
size and resolution to model the formation and evolution of galaxies
of a wide range of masses, and also include a counterpart set of dark
matter-only simulations of these volumes. The galaxy formation
simulations include the correct proportion of baryons and model gas
hydrodynamics and radiative cooling. State-of-the-art subgrid models
are used to follow star formation and feedback processes from both
stars and AGN.  The parameters of the subgrid model have been
calibrated to match certain observables as detailed in
\cite{Schaye2014}.  In particular, the simulations reproduce the
observed present day stellar mass function, galaxy sizes and many
other properties of galaxies and the intergalactic medium remarkably
well.  These simulations also show the correct trends with redshift of
many galaxy properties \citep{Schaye2014, Furlong2014}.

The simulations were run using an extensively modified version of the
code \gadget-3 \citep{Springel2008}, which is essentially a more
computationally efficient version of the public code \gadget-2
described in detail by \cite{Springel2005}. \gadget uses a Tree-PM
method to compute the gravitational forces between the \nbody
particles and implements the equations of hydrodynamics using Smooth
Particle Hydrodynamics \citep[SPH,][]{Monaghan1992, Price2010}.

The \eagle version of \gadget-3 uses an SPH implementation called
\anarchy (Dalla Vecchia in prep.), which is based on the general
formalism described by \cite{Hopkins2012}, with improvements to the
kernel functions \citep{Dehnen2012} and viscosity terms
\citep{Cullen2010}.  This new implementation of SPH alleviates some of
the problems associated with modelling contact
discontinuities and fluid instabilities.  As discussed by 
Dalla Vecchia (in prep.), the new formalism improves on the treatment of
instabilities associated with cold flows and filaments and on the 
evolution of the entropy of hot gas in halos.  The timestep limiter
of \cite{Durier2012} is applied to ensure good energy conservation
everywhere, including regions disturbed by violent feedback due to
supernovae and AGN.  The impact of this new hydrodynamics scheme on
our galaxy formation model is discussed by Schaller et al. (in prep.).

The analysis in this paper focusses on two simulations: the
Ref-L100N1504 simulation introduced by \cite{Schaye2014}, which is the
largest \eagle simulation run to date, and its counterpart dark
matter-only simulation, DM-L100N1504. To investigate smaller mass
halos and test for convergence in our results we also analyse the
higher resolution Recal-L025N0752 simulation (and its dark
matter-only counterpart) in which some of the sub-grid physics
parameters were adjusted to ensure that this calculation also
reproduces the observed galaxy stellar mass function, particularly at
the low-mass end, as discussed by \citep{Schaye2014}. We will refer to
the two simulations with baryon physics as ''\eagle'' simulations and
to the ones involving only dark matter as ''\dmonly'' simulations.

The main \eagle simulation models a cubic volume of side-length
$100~\rm{Mpc}$ with $1504^3$ gas and $1504^3$ dark matter particles to
redshift $z=0$. A detailed description of the initial conditions is
given in \cite{Schaye2014}. Briefly, the starting redshift was
$z=127$; the initial displacements and velocities were calculated
using second order Lagrangian perturbation theory with the method of
\cite{Jenkins2010}; the linear phases were taken from the public
multiscale Gaussian white noise field, Panphasia \citep{Jenkins2013};
the cosmological parameters were set to the best fit \lcdm values
given by the \emph{Planck-1} data \citep{Planck2013}:
$\left[\Omega_{\rm{m}}, \Omega_{\rm{b}},
  \Omega_\Lambda,h,\sigma_8,n_s\right] = \left[0.307, 0.04825, 0.693,
  0.6777, 0.8288, 0.9611\right]$; and the primordial mass fraction of
$\mathrm{He}$ was set to $0.248$.  These choices lead to a dark matter
particle mass of $9.70\times10^6\msun$ and an initial gas particle
mass of $1.81\times10^6\msun$. We use a comoving softening of
$2.66~\rm{kpc}$ at early times, which freezes at a maximum physical
value of $700~\rm{pc}$ at $z=2.8$. The Recal-L025N0752 simulation
follows $752^3$ gas and $752^3$ DM particles in a $25~\rm{Mpc}$ volume
assuming the same cosmological parameters. This implies a DM particle
mass of $1.21\times10^6\msun$ and an initial gas mass of
$2.26\times10^5\msun$. The softening is $1.33~\rm{kpc}$ initially and
reaches a maximum physical size of $350~\rm{pc}$ at $z=0$.

The \dmonly simulations, DM-L100N1504 and DM-L025N0752, follow
exactly the same volume as \eagle, but with only $1504^3$ and $752^3$
collisionless dark matter particles, each of mass
$1.15\times10^7\msun$ and $1.44\times10^6\msun$, respectively.  All
other cosmological and numerical parameters are the same as in the
\eagle simulation.

%-------------------------------------------------------------------------------
\subsection{Baryonic physics}
\label{ssec:baryonic_physics}

The baryon physics in our simulation correspond to the \emph{Ref}
\eagle model. The model, fully described in \cite{Schaye2014}, is
summarized here for completeness.

Star formation is implemented following \cite{Schaye2008}. A
polytropic equation of state, $P\propto \rho^{4/3}$, sets a lower 
limit to the gas pressure.  The star formation rate per unit mass
of these particles is computed using the gas pressure using an
analytical formula designed to reproduce the observed
Kennicutt-Schmidt law \citep{Kennicutt1998} in disk galaxies \citep{Schaye2008}. 
Gas
particles are converted into stars stochastically. The threshold in
hydrogen density required to form stars is metallicity dependent with
lower metallicity gas having a higher threshold, thus capturing the
metallicity dependence of the $\rm{HI}-\rm{H}_2$ phase transition 
\citep{Schaye2004}.

The stellar initial mass function is assumed to be that of
\cite{Chabrier2003} in the range $0.1\msun$ to $100\msun$ with each
particle representing a single age stellar population. After
$3\times10^7~\rm{yrs}$ all stars with an initial mass above $6\msun$
are assumed to explode as supernovae. The energy from these explosions
is transferred as heat to the surrounding gas. The temperature of an
appropriate amount of surrounding gas is raised instantly by
$10^{7.5}~\rm{K}$.  This heating is implemented stochastically on one
or more gas particles in the neighbourhood of the explosion site
\citep{DallaVecchia2012}. This gas, once heated, remains coupled in a
hydrodynamic sense with its SPH neighbours in the ISM, and therefore
exerts a form of feedback locally that can affect future star
formation and radiative cooling.

The energy injected into the gas corresponds to $10^{51}~\rm{erg}$ per
supernovae times a dimensionless efficiency factor, $f_{\rm E}$,
that depends on the local gas metallicity and density. The
construction of $f_{\rm E}$ and its impact on galaxy formation is discussed
thoroughly by \cite{Schaye2014} and \cite{Crain2014}. For a gas of metallicity, 
$Z$, and hydrogen
number density, $n_{\rm{H}}$, the efficiency in the reference model
is: 

\begin{equation}
 f_{\rm E} = 0.3 + 2.7 S\left(X;w\right),\nonumber
\end{equation}
where $w = 2/\ln10$,  
\begin{equation}
  X = 3.35\left(\frac{Z}{0.1Z_\odot}\right)\left( 
\frac{0.1~\rm{cm}^{-3}}{n_{\rm{H}}}
 \right),\nonumber
\end{equation}
and $S(X;w)$ is a convenient sigmoid function which varies between
0 and 1, and which we will need again in the following section.
We define the sigmoid function for $x\geq0$, $w>0$ as
\begin{equation}
 S(X;w) =  \frac{X^w}{1+X^w}.
\label{sigmoid}
\end{equation}
As $X$ varies from zero to infinity, the sigmoid function $S(X;w)$
smoothly varies between 0 and 1, taking the value of
$\frac{1}{2}$ when the argument $X=1$. The parameter $w$ controls the
rapidity of the transition between the asymptotes.

Besides energy from star formation, the star particles also
release metals into the ISM through three evolutionary channels: type
Ia supernovae, winds and supernovae from massive stars, and AGB
stars using the method discussed in \cite{Wiersma2009b}. The yields for each 
process are taken from
\cite{Portinari1998}, \cite{Marigo2001} and \cite{Thielemann2003}. Following 
\cite{Wiersma2009a}, the abundances
of the eleven elements that dominate the cooling rates are tracked.
These are used to compute element-by-element dependent cooling rates
in the presence of the Cosmic Microwave Background and the
ultraviolet and X-ray backgrounds from galaxies and quasars according
to the model of \cite{Haardt2001}.

For halos whose masses first exceed $M_{\rm{FOF}} = 10^{10}
h^{-1}\msun$ ($\approx1500$ dark matter particles, see section
\ref{ssec:halo_definition}), black hole (BH) sink particles are placed
at the centre of the halos.  The BHs are then allowed to grow through
gas accretion and by merging with other BHs using methods based on
those introduced by \cite{Springel2005b} and \cite{Booth2009}.  The
gas surrounding a BH is accreted at a rate given by the Bondi-Hoyle
formula \citep{Bondi1944} unless the viscous timescale of the gas
around the BH is larger than the Bondi time, in which case the
accretion rate is reduced by a factor proportional to the cube of the
ratio of the local sound speed and the rotation velocity
\citep{RosasGuevara2013}. For a BH of mass, $M_{\rm{BH}}$, surrounded
by gas at density, $\rho$, velocity with respect to the BH, $v$, and
sound speed, $c_{\rm{s}}$, the accretion rate is:
\begin{equation}
 \dot m_{\rm{BH}} = \frac{4\pi G 
M_{\rm{BH}}^2\rho}{\left(c_{\rm{s}}^2+v^2\right)^{3/2}}\cdot\left\{ 
  \begin{array}{ccl}
   \frac{1}{C_{\rm{visc}}}\left(\frac{c_{\rm{s}}}{V_\phi}\right)^3& \rm{if} & 
C_{\rm{visc}}V_\phi^3>c_{\rm{s}}^3\\
   1& \rm{if} & C_{\rm{visc}}V_\phi^3\leq c_{\rm{s}}^3
  \end{array}
\right., \nonumber
\end{equation}
 where $V_\phi$ is the circular speed of the gas at the Bondi radius
and $C_{\rm{visc}} = 2\pi$ in the reference simulation. 

Feedback due to AGN activity is implemented in a similar way to the
feedback from star formation described above.  The fraction of the
accreted rest mass energy liberated by accretion is 
$\epsilon_r=0.1$, and the heating efficiency of this liberated energy
(i.e. the fraction of the energy that couples to the gas phase) is
$\epsilon_f=0.15$. Gas particles receiving AGN feedback energy are
chosen stochastically and their temperature is raised by
$10^{8.5}~\rm{K}$.

These models of supernova and AGN feedback are extensions of the
models developed for the Virgo Consortium projects \textsc{owls}
\citep{Schaye2010} and \textsc{gimic} \citep{Crain2009}.  The values
of the parameters were constrained by matching key observables of the
galaxy population including the observed $z\approx0$ galaxy stellar
mass function, galaxy sizes and the relation between black hole and
stellar mass \citep{Crain2014}.

%-------------------------------------------------------------------------------

\subsection{Halo definition and selection}
\label{ssec:halo_definition}

Halos were identified using the Friends-of-Friends (FOF) algorithm on
all dark matter particles adopting a dimensionless linking length,
$b=0.2$ \citep{Davis1985}. We then applied the \subfind algorithm,
which is built into \gadget-3 \citep{Springel2001, Dolag2009}, to
split the FOF groups into self-bound substructures. A sphere is grown
outwards from the potential minimum of the dominant subgroup out to a
radius where the mean interior density equals a target value. This
target value is conventionally defined in units of the critical
density, $\rhocr(z)={3H^2(z)}/{8\pi G}$. With our choice of cosmology,
at $z=0$ we have $\rhocr = \rhocr(0) = 127.5~\msun~\rm{kpc}^{-3}$.  A
halo of mass, $M_{\rm{X}}$, is then defined as all the mass within the
radius, $R_{\rm{X}}$, for which
\begin{equation}
 \frac{3M_{\rm{X}}}{4\pi R_{\rm{X}}^3} = \rm{X}\rhocr(z)
\end{equation}
Commonly used values are $\rm{X}=200$, $500$ and $2500$,
leading to the definition of the mass, $M_{200}$, and the radius,
$R_{200}$, and similar definitions for other values of $\rm{X}$.

In the particular case of the virial radius, $R_{\rm vir}$, one can use the
spherical top-hat collapse model to derive the value of $\rm{X}$ 
\citep{Eke1996}. We use the
fitting formula given by \cite{Bryan1998}:
\begin{equation}
 \rm{X} = 18\pi^2+82\left(\Omega_{\rm{m}}(z) -1\right) - 
39\left(\Omega_{\rm{m}}(z) -1\right)^2,
\end{equation}
where
\begin{equation} 
\Omega_{\rm{m}}(z)=\Omega_{\rm{m}}\left(1+z\right)^3\left(\frac{H_0}{H(z)}
\right)^2,
\end{equation}
and $H(z)$ is the value of the Hubble parameter at redshift $z$ which, in a flat 
Universe, is
\begin{equation}
 H(z) = H_0\sqrt{\Omega_{\rm{m}}(1+z)^3+\Omega_\Lambda}.
\end{equation}
In the case of the \emph{Planck1} cosmology, at $z=0$, $\rm{X}=102.1$,
giving $M_{\rm{vir}} = M_{102}$ and $R_{\rm{vir}} = R_{102}$.

We define the circular velocity, $V_{\rm{X}}$, as
\begin{equation}
 V_{\rm{X}} = \sqrt{\frac{GM_{\rm{X}}}{R_{\rm{X}}}}.
\end{equation}
We only consider halos with more than $200$ particles within
$R_{200}$, implying a limit, $M_{200} \gtrsim 2.5\times10^8\msun$, in
our joint analysis of the two \eagle\ simulations. For specific properties
that depend on the internal structure of the halo we adopt more
conservative limits as described in section \ref{sec:HaloProfile}.

\subsection{Matching halos between the two simulations}
\label{ssec:halo_matching}

The \eagle and \dmonly simulations start from identical
Gaussian density fluctuations.  Even at $z=0$ it is possible, 
in most cases, to identify matches between halos in the two
simulations.  These matched halos are comprised of matter that
originates from the same spatial locations at high redshift in the two
simulations. In practice, these identifications are made by matching
the particle IDs in the two simulations, as the values of the IDs
encode the Lagrangian coordinates of the particles in the same way in
both simulations.

For every FOF group in the \eagle simulation, we select the 50 most bound
dark matter particles. We then locate those particles in the \dmonly
simulation. If more than half of them are found in a single FOF group
in the \dmonly simulation, we make a link between those two halos. We then
repeat the procedure by looping over FOF groups in the \dmonly
simulation and 
looking for the position of their counterparts in the \eagle simulation. More
than $95\%$ of the halos with $M_{200} > 2\times 10^{10}\msun$ can be
matched bijectively, with the fraction reaching unity for halos above
$7\times 10^{10}\msun$ in the L100N1504 volumes. Similarly, $95\%$ of
the halos with $M_{200} > 3\times10^9$ can be matched bijectively in
the L025N0752 volumes.

%###############################################################################
\section{Halo masses and content}
\label{sec:HaloCounts}

Previous work comparing the masses of halos in cosmological galaxy
formation simulations with matched halos in counterpart dark
matter-only simulations have found strong effects for all but the most
massive halos \citep[e.g.][]{Cui2012, Sawala2013}. \cite{Sawala2013}
found that baryonic effects can reduce the masses of halos by up to
$25\%$ for halo masses (in the dark matter only simulation) below
$10^{13}\msun$. (They did not include AGN feedback in their
simulation.) A similar trend was observed at even higher masses by
\cite{Martizzi2013}, \cite{Velliscig2014}, \cite{Cui2014} and
\cite{Cusworth2013} using a variety of subgrid models for star
formation and stellar and AGN feedback. All these authors stress that
their results depend on the details of the subgrid implementation
used.  This is most clearly shown in \cite{Velliscig2014}, where the
amplitude of this shift in mass is shown explicitly to depend on the
subgrid AGN feedback heating temperature, for example. Hence, it is
important to use simulations that have been calibrated to reproduce
the observed stellar mass function.

In this section we find that similar differences to those seen before
occur between halo masses in the \eagle and \dmonly models. These
differences are of particular interest because \eagle reproduces well
a range of low-redshift observables of the galaxy population
such as masses, sizes and star formation rates \citep{Schaye2014},
although the properties of clusters of galaxies are not reproduced as
well as in the Cosmo-\textsc{owls} simulation \citep{LeBrun2014} analyzed
by \cite{Velliscig2014}.

%-------------------------------------------------------------------------------

\subsection{The effect of baryon physics on the total halo mass}
\label{ssec:HaloMass}

In this section we compare the masses of halos in the \eagle and
\dmonly simulations combining our simulations at two different
resolutions.  To minimise any possible biases due to incomplete
matching between the simulations, we only consider halos above
$3\times10^{9}\msun$ (in \dmonly), since these can be matched
bijectively to their counterparts in more than $95\%$ of cases.

\begin{figure}
\includegraphics[width=\columnwidth]{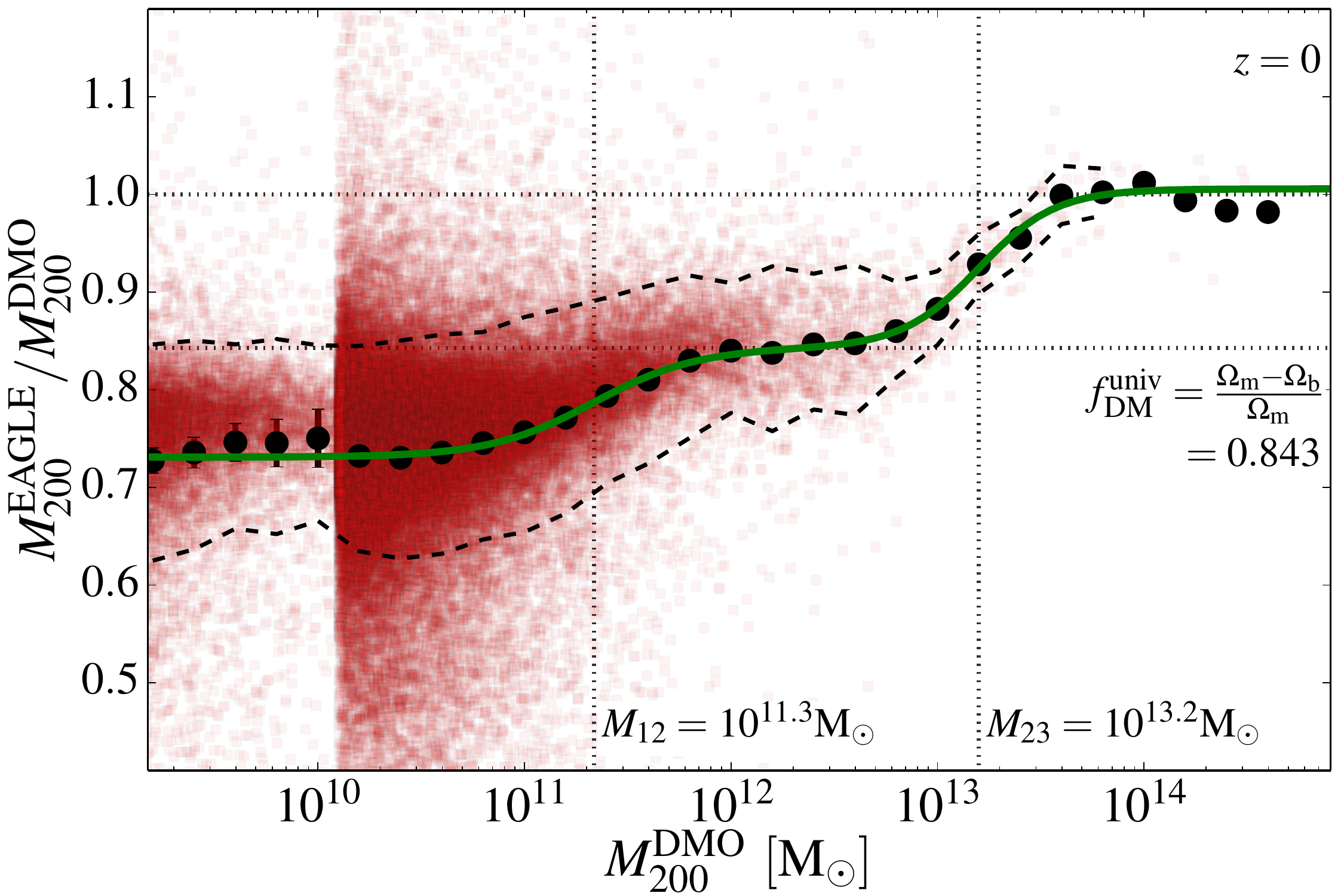}
\caption{The ratio of the masses of the matched halos in the \eagle
  and \dmonly simulations. The red squares show values for individual
  halos and the black filled circles values binned by \dmonly halo
  mass. Halos with $M_{200}^{\rm DMO} < 10^{10.1}\msun$ are extracted
   from the higher resolution, L025N0752, simulation.  The binned points are 
the geometric 
  average of the individual ratios with the error bars at
  $M_{200}^{\rm DMO} < 10^{10.1}\msun$ indicating the uncertainty
  arising from the low number of halos in the high-resolution
  simulation.  The black dashed lines placed above and below the black
  points show the geometrical $1\sigma$ scatter for each bin. The
  lower horizontal grey dotted line indicates the universal dark
  matter fraction $f_{\rm DM} = 1 - f_{\rm b} = (\Omega_{\rm{m}} -
  \Omega_{\rm{b}}) / \Omega_{\rm{m}} = 0.843$. The upper dotted line
  marks unity. The green solid line is the function of
  Eqn.~\ref{eq:meFit} fitted to the binned ratios. The vertical dotted
  lines mark the values of the fitting parameters $M_{12}$ and
  $M_{23}$.}
\label{fig:MatchedHalos}
\end{figure}

Fig.~\ref{fig:MatchedHalos} shows the ratio of $M_{200}$ for matched
halos in the \eagle and \dmonly simulations as a function of $M_{200}$
in the \dmonly simulation.  The black filled circles correspond to the
geometric mean of the ratios in each logarithmically spaced mass
bin. The choice of a geometric mean is motivated simply by the fact
that its reciprocal is the geometric mean of
$M_{200}^{\rm{DMO}}/M_{200}^{\rm{EAGLE}}$, which is also a quantity of
interest.

The halos in \eagle are typically lighter than their \dmonly
counterparts. There appear to be three distinct regimes in
Fig.~\ref{fig:MatchedHalos} . At the low mass end,
$M_{200}<5\times10^{10}~\msun$,
$M_{200}^{\rm{EAGLE}}/M_{200}^{\rm{DMO}}$ drops to $\sim0.72$.  This
is less than one minus the universal baryon fraction, $f_{\rm DM}$, so not
only have the baryons been removed but the dark matter has also been
disturbed. The reduction in mass due to the loss of baryons lowers the
value of $R_{200}$ and thus the value of $M_{200}$. However, this
reduction in radius is not the sole cause for the reduction in halo
mass: the amount of mass within a fixed physical radius is actually
lower in the simulation with baryons because the loss of baryons,
which occurs early on, reduces the growth rate of the halo
\citep{Sawala2013}. At higher masses, stellar feedback becomes less
effective, but AGN feedback can still expel baryons and the ratio
rises to a plateau of $\sim0.85$ between
$M_{200}^{\rm{DMO}}=10^{12}~\msun$ and $5\times10^{12}~\msun$.
Finally, for the most massive halos ($M_{200} > 10^{14}~\msun$) not
even AGN feedback can eject significant amounts of baryons from the
halos and the mass ratio asymptotes to unity.

\cite{Sawala2013} proposed a fitting function to the ratio of
$M_{200}$ in simulations with and without baryons from the
\textsc{gimic} project \citep{Crain2009}.  Their study focused mostly
on lower-mass objects and subhalos, but included enough large halos to
sample the high-mass end of the relation. Their four parameter fitting
function can be written as:
\begin{equation}
 \frac{M_{200}}{M_{200}^{\rm{DMO}}} = a + 
(b-a)S\left(\frac{M_{200}^{\rm{DMO}}}{M_t};w\right),
\label{eq:tillFit}
\end{equation}
where $S$ is a sigmoid function that varies smoothly between 0 and 1, 
and is defined in Eqn.~\ref{sigmoid}. The best-fit parameter values in 
\cite{Sawala2013} are:
$(a,b,\log_{10}(M_t/\msun),w)$ = $(0.69,0.98,11.6,0.79)$.  The values of $a$
and $b$ correspond to the low- and high-mass asymptotes, respectively.

\cite{Velliscig2014} used a similar fitting function to summarise the
results of their study, again with four parameters, which can be
written as:
\begin{equation}
 \frac{M_{200}}{M_{200}^{\rm{DMO}}} = a\left( \frac{b}{a} 
\right)^{S\left(M_{200}^{\rm{DMO}}/M_t;w\right)},
 \label{eq:marcoFit}
\end{equation}
where exactly the same sigmoid function is used to interpolate between the two
asymptotic values, $a$ and $b$, but now in a geometric rather
than arithmetic fashion.   The functional forms of
Eqns.~\ref{eq:tillFit} and \ref{eq:marcoFit} are virtually
identical as, in practice, the ratio $b/a$ is never far from unity.

It is quite clear, however, from Fig.~\ref{fig:MatchedHalos} that a
single sigmoid function does not reproduce the behaviour we observe
particularly well: the ratio shows three, not two, distinct plateaux.
The simulations used by \cite{Sawala2013} did not include AGN feedback
and so did not show the change in mass arising from this form of
feedback. In contrast, the simulations used by \cite{Velliscig2014}
did not have sufficient numerical resolution to see the asymptotic
low-mass behaviour determined by stellar feedback.

To fit our results, we use a double sigmoid:
\begin{eqnarray}
 \frac{M_{200}}{M_{200}^{\rm{DMO}}} = r_1 & + & 
(r_2-r_1)S\left(\frac{M_{200}^{\rm{DMO}}}{M_{12}};t_{12}\right)
 \nonumber\\
                                                 & + & 
(r_3-r_2)S\left(\frac{M_{200}^{\rm{DMO}}}{M_{23}};t_{23}\right),
\label{eq:meFit}
\end{eqnarray}
where the seven parameters can be interpreted as follows: $r_1$, $r_2$
and $r_3$ are the values of the ratios corresponding to the three
distinct plateaux; the mass scales, $M_{12}$ and $M_{23}$, are the
mid-points between regimes 1 and 2, and 2 and 3 respectively;
and the parameters, $t_{12}$ and $t_{23}$, control the rapidity of
each transition.

The green curve in Fig.~\ref{fig:MatchedHalos} shows the best fitting
curve to the black binned data points.  The fit was obtained by a
least-squares minimisation for all seven parameters assuming Poisson
uncertainties for each mass bin. Adopting a constant error instead
gives very similar values for all parameters.  The values of the two
transition masses, $M_{12}$ and $M_{23}$, are shown as vertical dotted
lines in Fig.~\ref{fig:MatchedHalos}. The best-fitting parameters are
given in Table \ref{tab:bestFit}. Note that the value of $r_3$ is, as
expected, very close to unity.

\begin{table}
  \caption{Best fitting parameters to the
    black points in Fig.~\ref{fig:MatchedHalos} using Eqn.~\ref{eq:meFit},
    and their uncertainties which are taken to be the diagonal elements
    of the correlation matrix of the least-squares fitting  procedure.}
\label{tab:bestFit}
\begin{center}
\begin{tabular}{|c|r|r|}
Parameter & Value & $1\textendash\sigma$ fit uncertainty\\
\hline
 $r_1$ & $0.7309                $&$\pm 0.0014 $ \\
 $r_2$ & $0.8432                 $&$\pm 0.0084 $ \\
 $r_3$ & $1.0057                 $&$\pm 0.0024 $ \\
 $\log_{10}(M_{12}/\msun)$ & $11.33  $&$\pm 0.003 $ \\
 $\log_{10}(M_{23}/\msun)$ & $13.19  $&$\pm 0.029 $ \\
 $t_{12}$ & $1.721              $&$\pm 0.045 $ \\
 $t_{23}$ & $2.377            $&$\pm 0.18 $ \\
\hline
\end{tabular}
\end{center}
\end{table}

The value of the first transition mass, $M_{12}=10^{11.35}\msun$, is
similar to that reported by \cite{Sawala2013} who found
$M_t=10^{11.6}\msun$ for the \textsc{gimic} simulations.  The second
transition, $M_{32}=10^{13.2}\msun$, is located well below the range
of values found by \cite{Velliscig2014} ($10^{13.7}\msun$
-$10^{14.25}\msun$).  However, as \cite{Schaye2014} have shown the AGN
feedback in the few rich clusters formed in the \eagle volume may not
be strong enough, as evidenced by the fact that this simulation
overestimates the gas fractions in clusters, whereas the
$400~\rm{Mpc}/h$ Cosmo-\textsc{owls} simulation used by
\cite{Velliscig2014} reproduces these observations
\citep{LeBrun2014}.

A simulation with stronger AGN feedback, \eagle-AGNdT9, which gives a
better match to the group gas fractions and X-ray luminosities than
\eagle, was discussed by \cite{Schaye2014} . Applying the same halo
matching procedure to  this simulation and its collisionless dark
matter-only counterpart, we obtain slightly different values for the
best-fitting parameters of Eqn.~\ref{eq:meFit}. The difference is
mainly in the parameters, $M_{23}$ and $t_{23}$, which describe the
high-mass end of the double-sigmoid function. In this model, the
transition occurs at
$\log_{10}\left(M_{23}/\msun\right)=13.55\pm0.09$, closer to the
values found by \cite{Velliscig2014}.  The width of the transition,
however, is poorly constrained, $t_{23}=3.0\pm12.7$, due to the small
number of halos (only eight with
$M_{200,\rm{DMO}}>2\times10^{13}\msun$) in this simulation which had
only an eighth the volume of the reference simulation.

As \cite{Velliscig2014} did, we provide a fit to the scatter in the
log of the ratio about the mean relation, valid over the range where
appropriately constraining data are available:
\begin{equation}
\sigma\left(\log_{10}(M_{200}^{\rm{DMO}})\right) = 0.044 - 0.015 \log_{10}
 \left(\frac{M_{200}^{\rm{DMO}}}{10^{12}\msun}\right).
 \label{eq:Myfitscatter}
\end{equation}
The scatter is about 10\% for a halo mass of $10^{12} \msun$ and
decreases with mass. The slope in the relation is
approximatively a factor of two greater than that found for the AGN models
of \cite{Velliscig2014}.

%-------------------------------------------------------------------------------

\subsection{The halo mass function}
\label{ssec:HMF}

\begin{figure}
\includegraphics[width=\columnwidth]{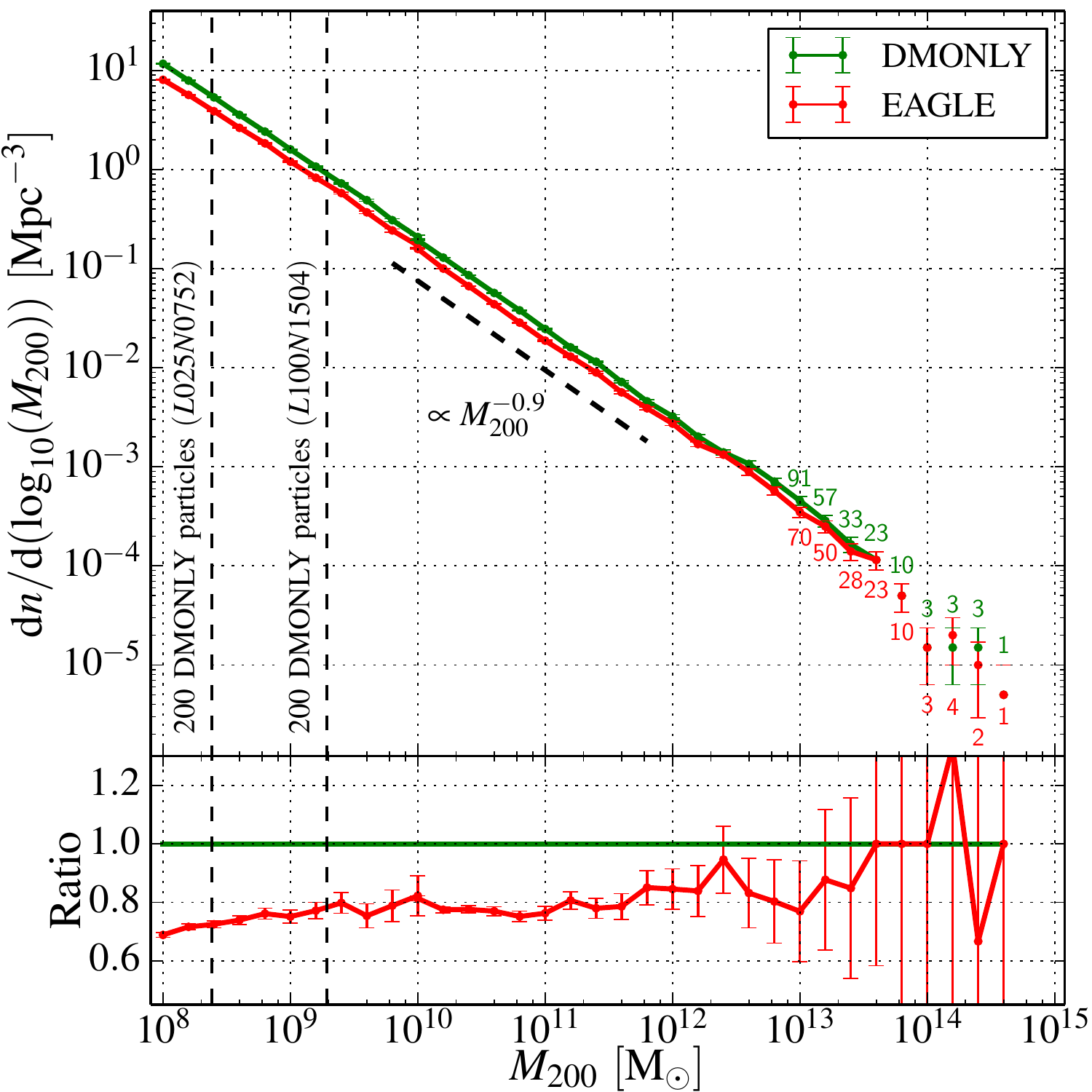}
\caption{Top panel: the abundance of halos at $z=0$ as a function of
  the mass, $M_{200}$, in the \eagle (red curve, lower line) and
\dmonly (green curve, upper line) simulations. The high resolution volume is 
used for $M_{200}^{\rm DMO} < 
10^{10.1}\msun$. The resolution limits for both simulations are indicated by
the  vertical dashed lines on the left,  and the number of halos in sparsely
populated bins is given above the Poisson error bars.
Bottom panel: the ratio of the mass functions in the \eagle and \dmonly 
simulations.}
\label{fig:HMF}
\end{figure}

The effect of baryons on the halo mass function can be seen in
Fig.~\ref{fig:HMF}.  The red and green lines in the top panel show the
mass functions in the \eagle and \dmonly simulations.  The ratio of
the two functions (bottom panel) shows an almost constant shift over
most of the plotted mass range, $M_{200} / \msun = 10^9 -
10^{13}$, as expected from Fig.~\ref{fig:MatchedHalos}. The relatively small 
volume of the \eagle simulation does
not sample the knee of the halo mass function well, but extrapolating
the fit to the mass ratios of Eqn.~\ref{eq:meFit} to higher masses,
together with results from previous studies \citep{Cusworth2013,
  Martizzi2013, Velliscig2014}, suggests that the differences vanish
for the most massive objects. Studies that rely on galaxy clusters to
infer cosmological parameters will need to take account of the effects
of the baryons, particularly for clusters of mass $M_{200} \lesssim
10^{14}\msun$.

%-------------------------------------------------------------------------------

\subsection{Baryonic and stellar fractions in the {\bf{\sc eagle}}
  simulation}
\label{ssec:baryon_fraction}

We have shown in the previous subsection that for all but the most
massive examples, halo masses are systematically lower when baryonic
processes are included.  In this subsection we examine the baryonic
content of halos in the \eagle simulation. We restrict our analysis to
the L100N1504 volume.

\begin{figure}
\includegraphics[width=\columnwidth]{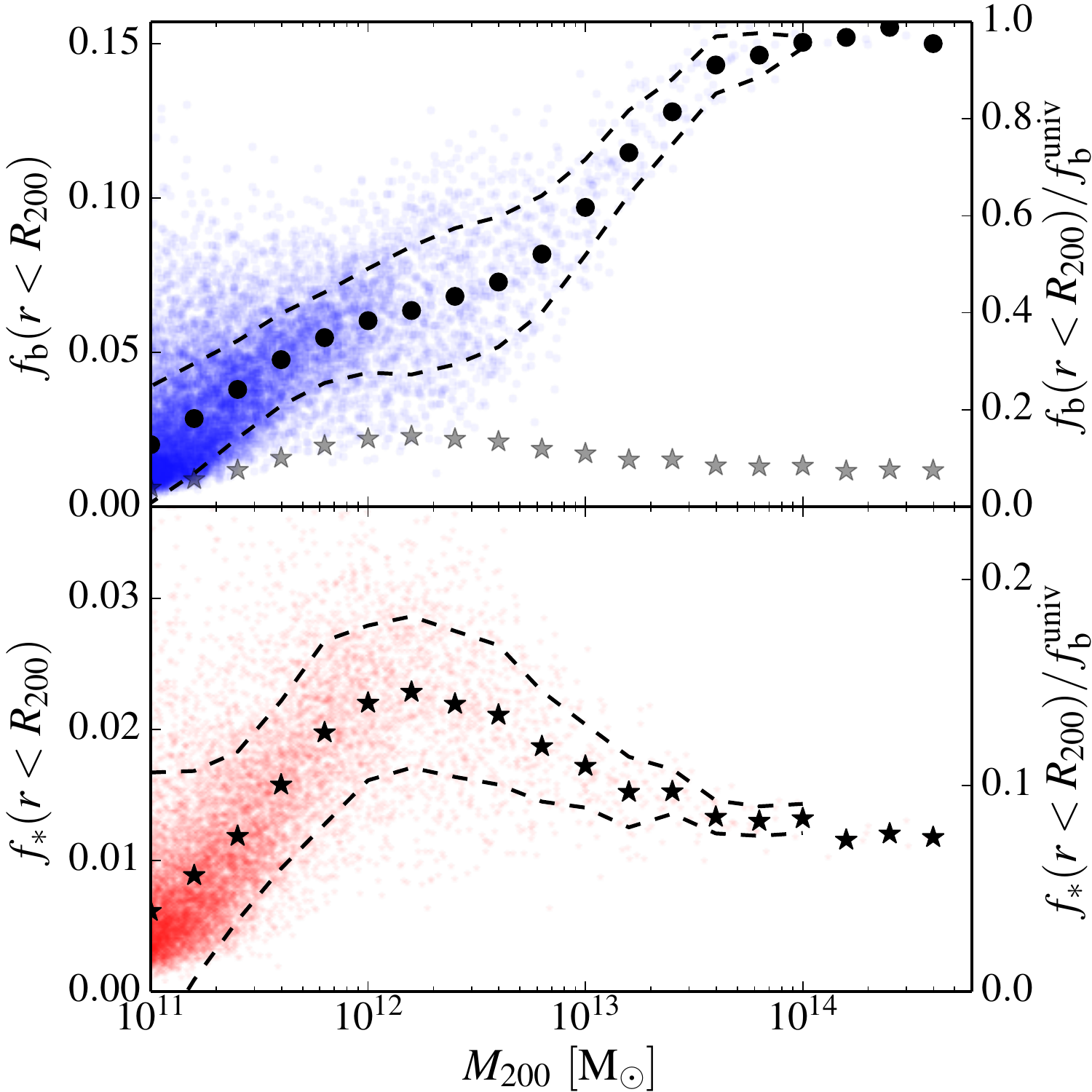}
\caption{Baryon fraction, $f_{\rm b}=M_{\rm b}/M_{200}$ (top panel), and stellar
  fraction, $f_*=M_*/M_{200}$ (bottom panel), within $R_{200}$ as a
  function of $M_{200}$.  The right-hand axis gives the fractions in
  units of the universal mean value, $f_{\rm b}^{\rm{univ}}=0.157$. The
  solid circles in the top panel and the stars in the bottom panel
  show the mean value of the fractions binned by mass. The dashed
  lines above and below these symbols show the \rms\ width of each bin
  with more than three objects.  The stellar fractions are reproduced
  as grey stars in the top panel.}
\label{fig:stellarFraction}
\end{figure}

Fig.~\ref{fig:stellarFraction} shows the mass fractions of baryons and
stars within $R_{200}$ as a function of the halo mass, $M_{200}$, in
the \eagle simulation.  The baryon fraction increases with halo mass
and approaches the universal mean value, $f_{\rm b}^{\rm{univ}} \equiv
\Omega_{\rm{b}}/\Omega_{\rm{m}}$, for cluster mass halos. The gas is
the most important baryonic component in terms of mass over the entire
halo mass range.  At a much lower amplitude everywhere, the stellar
mass fraction peaks around a halo mass scale of $2\times10^{12}\msun$
where star formation is at its least inefficient.

The baryon fractions are much lower than the universal value for all but the 
most
massive halos. For Milky Way sized halos, we find $f_{\rm b} /
f_{\rm b}^{\rm{univ}} \approx 0.35$. It is only for group and cluster sized
halos, whose deeper gravitational potentials are able to retain most of
the baryons even in the presence of powerful AGN, that the baryon
fraction is close to $f_{\rm b}^{\rm{univ}}$. The baryon fractions of the
halos extracted from the \eagle-AGNdT9 model (which provides a better
match to X-ray luminosities; \citealt{Schaye2014}) are presented in
Appendix~\ref{ssec:AGN_changes}.

The stellar mass fraction is never more than a few percent. At the
peak, around $M_{200}\approx2\times10^{12}\msun$, it reaches a value
of $\sim0.023$. Multiplying the stellar fraction by the halo mass
function leads to an approximate stellar mass function, which is close
to the actual one (published in \citealt{Schaye2014}), after a fixed
aperture correction is applied to mimic observational measurements.
As may be seen in both panels, there is significant scatter in the
baryonic and stellar fractions, with variations of a factor of a few
possible for individual halos.

\begin{figure}
\includegraphics[width=\columnwidth]{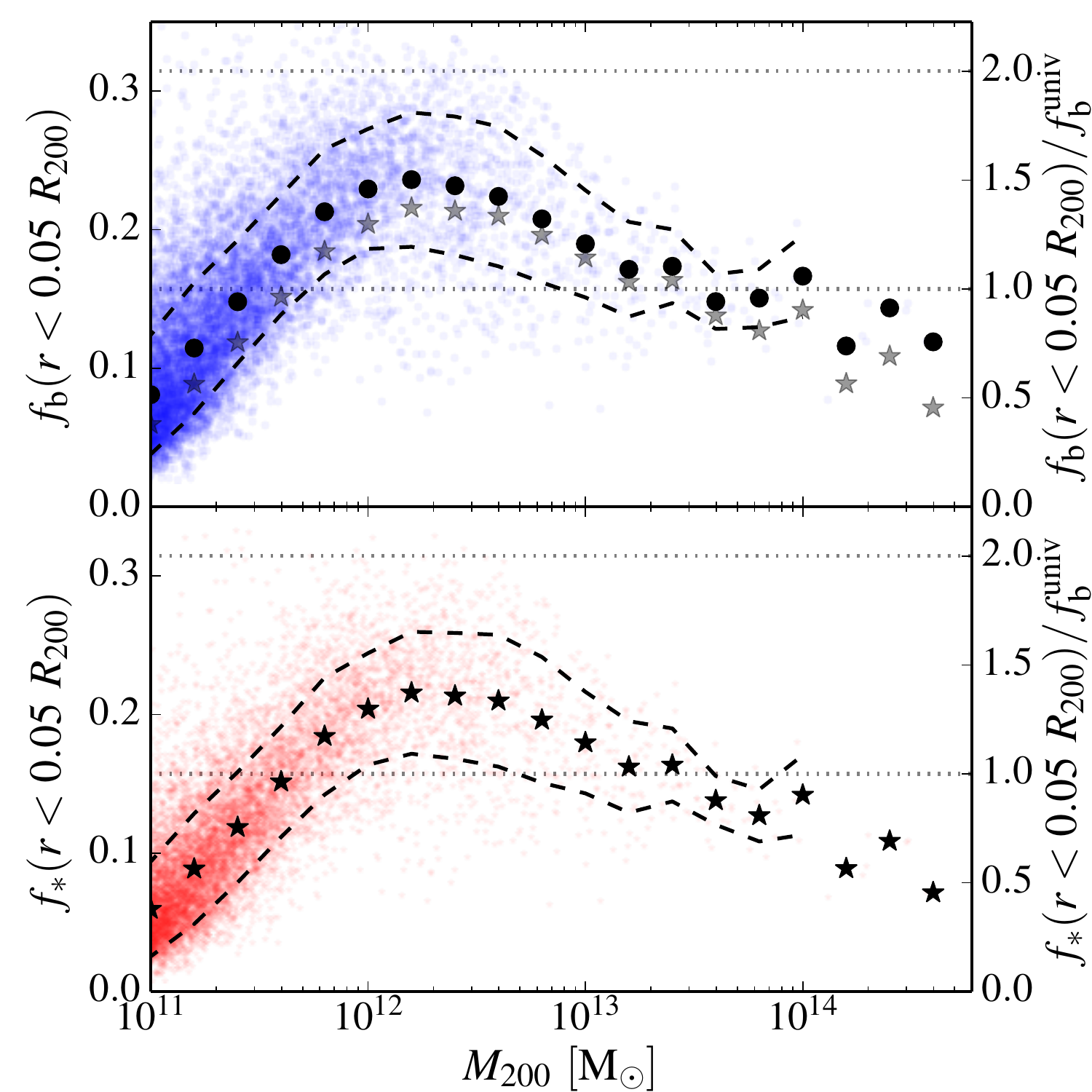}
\caption{Same as Fig.~\ref{fig:stellarFraction} but for the mass
  contained within 5\% of $R_{200}$. Note the different scale on the
  ordinate axis. The dotted horizontal lines mark one and two times
  the universal baryon fraction.}
\label{fig:stellarFractionCentre}
\end{figure}

While the baryonic and stellar fractions are low within $R_{200}$,
they are much higher in the inner regions of halos as shown in
Fig.~\ref{fig:stellarFractionCentre}, where these fractions are now
plotted within $0.05R_{200}$, a scale commensurate with the sizes of
galaxies both in \eagle and in the real universe. Within this radius
the fractions rise above the cosmic mean for halos in the mass range
$5\times10^{11}\msun<M_{200}<2\times10^{13}\msun$.  The central parts
of these halos are strongly dominated by the baryons.  In agreement
with observations of the nearby universe, the most important
contribution to the mass on these scales is from stars rather than
gas. Another notable feature is that the most massive halos are
baryon poor in their central regions, reflecting the regulation by AGN
feedback.

\section{ Halo profiles}
\label{sec:HaloProfile}

In this section we explore the effects of baryons on halo profiles
restricting the analysis to halos with more than $5000$ particles
within $R_{\rm vir}$, which corresponds to a halo mass of about
$5\times10^{10} \msun$ in the L100N1504 simulation and $6\times10^{9}
\msun$ in the L050N0752 simulation. The stellar masses found in the
\eagle simulation for halos of this mass are consistent with
observational expectations based on abundance matching
\citep{Schaye2014}. Halos smaller than this typically have fewer than
the hundred star particles, which \cite{Schaye2014} showed to be a
necessary criterion for many applications.  This limit of 5000 in the
number of particles is intermediate between those used in other
studies.  It is similar to the number adopted by \cite{Ludlow2013} and
lower than the number adopted by \cite{Neto2007} and
\cite{Duffy2008,Duffy2010} ($10000$ particles), but higher than the
number adopted by \cite{Gao2008,Dutton2014} ($3000$ particles) or
\cite{Maccio2007} ($250$ particles).  There are $22867$ halos with at
least $5000$ particles in the Ref-L100N1504 \eagle simulation and
$2460$ in the Recal-L025N0752 simulation.

We define \emph{relaxed} halos as those where the separation between
the centre of the potential and the centre of mass is less than
$0.07R_{\rm vir}$, as proposed by \cite{Maccio2007}. \cite{Neto2007}
used this criterion, and also imposed limits on the substructure
abundance and virial ratio. \cite{Neto2007} found that the first
criterion was responsible for rejecting the vast majority of unrelaxed
halos. Their next most discriminating criterion was the amount of mass
in substructures. In common with \cite{Gao2008}, here we use stacked
profiles. Hence, individual substructures, which can be important when
fitting individual halos, have a smaller effect on the average
profile. We therefore do not use a substructure criterion to reject
halos. Our relaxed sample includes $13426$ halos in the L100N1504
simulation and $1590$ in the L025N0752 simulation.  We construct the
stacked halos by coadding halos in a set of contiguous bins of width
$\Delta \log_{10}(M_{200}) = 0.2$.

The density and mass profiles of each halo and of the stacked halos
are obtained using the procedure described by \cite{Neto2007}. We
define a set of concentric contiguous logarithmically spaced spherical
shells of width $\Delta\log_{10}(r)=0.078$, with the outermost bin
touching the virial radius, $R_{\rm vir}$. The sum of the masses of
the particles in each bin is then computed for each component (dark
matter, gas, stars, black holes) and the density is obtained by
dividing each sum by the volume of the shell.

\subsection{Resolution and convergence considerations}
\label{ssec:resolution_test}

Determining the minimum radius above which the results are robust and
reliable is non-trivial. For DM-only simulations, \cite{Gao2008}
showed that the best fit NFW profiles are sensitive to this choice and
it is, therefore, important to estimate this minimum converged radius
accurately. For DM-only simulations the thorough resolution study of
\cite[][P03]{Power2003}  suggests a convergence radius, $R_{P03}$, based
on the two-body relaxation timescale of particles orbiting in the
gravitational potential well. This criterion can be written
as:
\begin{equation} 0.6 \leq  
\frac{\sqrt{200}}{8}\sqrt{\frac{4\pi\rhocr}{3m_{\rm{DM}}}}\frac{\sqrt{N(<R_{P03}
)}}{\ln
    N(<R_{P03})}R_{P03}^{3/2},
\label{eq:P03}
\end{equation}
where $N(<r)$ is the number of particles of mass, $m_{\rm{DM}}$,
within radius $r$.

While this criterion could be applied to the \dmonly simulation, the
situation for the \eagle simulation is more complex since, as
discussed by \cite{Schaye2014}, the concept of numerical convergence
for the adopted subgrid model is itself ill defined. One option would
be simply to apply the P03 criterion, which is appropriate for the
\dmonly simulation, to both simulations.  Alternatively, we could
apply the criterion to the dark matter component of the halos in the
baryon simulation or to all the collisionless species (stars, dark
matter and black holes).  Neither of these options is fully
satisfactory but, in practice, they lead to similar estimates for
$R_{P03}$. For the smallest halos of the L100N1504 simulation considered in
this section, we find $R_{P03} \approx 5.1~\rm{kpc}$ whereas for the
largest clusters we obtain $R_{P03} \approx 3.5~\rm{kpc}$.

\begin{figure*}
\includegraphics[width=\textwidth]{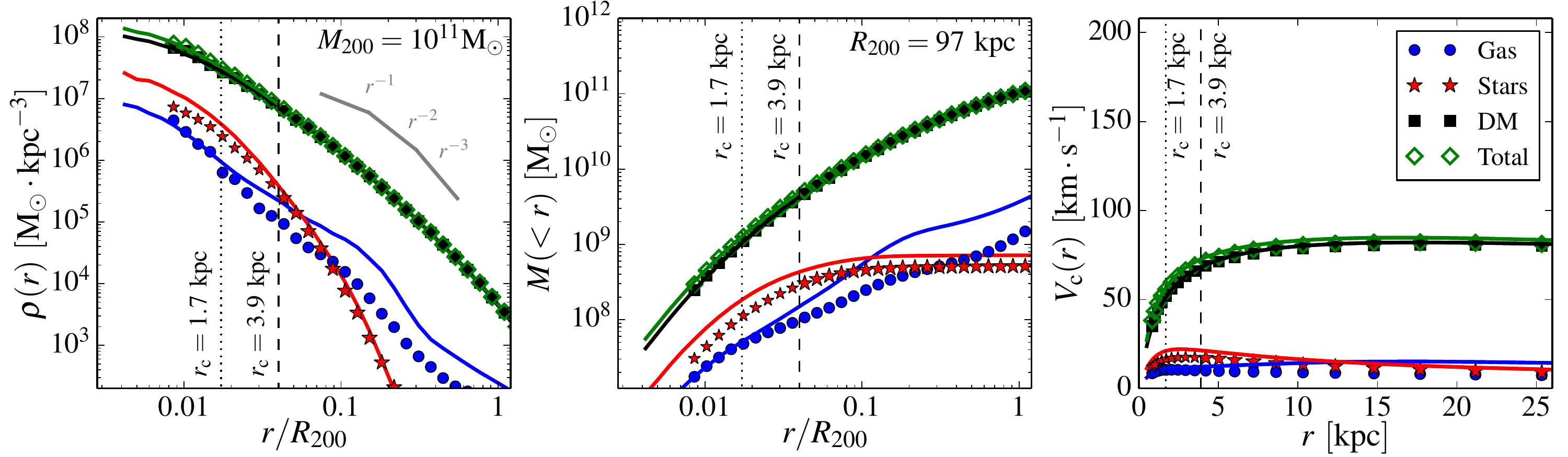}
\caption{From left to right: the density, mass and circular velocity
  profiles of a stack of the 44 relaxed halos of mass $10^{11}\msun$
  at $z=0$ that are present in both the L025N0752 simulation (lines)
  and the L025N0376 simulation (symbols). Profiles of total
  matter (green), dark matter (black), gas (blue) and the stellar
  component (red) are shown for both resolutions.  The vertical dashed
  and dotted lines show the resolution limits, $r_{\rm{c}}$, derived
  from our modified P03 criterion for the L025N0376 and L025N0752
  simulations respectively; data point are only shown at radii larger
  than the Plummer equivalent force softening. The dark matter, total
  matter and stellar profiles are well converged even at radii smaller
  than $r_{\rm c}$, indicating that this convergence cirterion is very 
  conservative when relaxed halos in a narrow mass range are averaged
  together. Convergence is much poorer for the subdominant gas
  distribution at large radii.}
\label{fig:profilesResolution}
\end{figure*}

The original P03 criterion ensures that the mean density internal to
the convergence radius, $\bar\rho = 3M(r<R_{P03}) / 4\pi R_{P03}^3$,
is within $10\%$ of the converged value obtained in a simulation of
much higher resolution. As the magnitude of the differences between
the \eagle and \dmonly profiles that we see are significantly larger
than 10\% typically, we can relax the P03 criterion somewhat.
Reanalysing their data, we set the coefficient on the left-hand side
of Eqn.~\ref{eq:P03} to $0.33$, which ensures a converged value of the
mean interior density at the $20\%$ level. With this definition, our
minimal convergence radius $r_{\rm{c}}$ takes values between
$4~\rm{kpc}$ and $2.9~\rm{kpc}$ for halos with $M_{200} \sim
10^{11}\msun$ up to $M_{200}\sim 10^{14}\msun$.  Similarly, in the
L025N0752 simulation our modified criterion gives $r_{\rm
  c}\approx1.8~\rm{kpc}$. Note that despite adopting a less
conservative criterion than P03, the values of $r_{\rm{c}}$ are always
greater than the Plummer equivalent softening length where the force
law becomes Newtonian, $2.8\epsilon = 0.7~\rm{kpc}$ in the L100N1504
simulation and $0.35~\rm{kpc}$ in L025N0752 simulation.

The validity of our adopted convergence criterion can be tested
directly by comparing results from our simulations at two different
resolutions. Specifically, we compare our two simulations of $(25~
\rm{Mpc})^3$ volumes, L025N0752, and L025N0376 which has the same
initial phases as L025N0752 but the resolution of the reference,
L100N1504, simulation.  In the language of \cite{Schaye2014}, this is
a \emph{weak} convergence test since the parameters of the subgrid
models have been recalibrated when increasing the resolution.

Fig. ~\ref{fig:profilesResolution} shows the stacked profiles of the
$44$ relaxed halos of mass $10^{11}\msun$ present in both the
L025N0376 and L025N0752 simulations. This mass bin contains enough
halos for the stacks not to be dominated by Poisson noise and the
halos are large enough to contain more than $5000$ particles in the
lower resolution simulation. The three panels show density, contained
mass and circular velocity profiles respectively, using symbols for
the default resolution and lines for the higher resolution
simulation. As may be seen, the stacked dark matter and total matter
profiles are very well converged over most of the radial range, both
in terms of the integral quantities, $M(r)$ and $V_{\rm c}(r)$, and in
terms of the differential quantity, $\rho(r)$. The dashed and dotted
vertical lines show the convergence radius, $r_{\rm c}$, for the
default and high resolution simulations respectively, computed
following the procedure described above.

The dark matter and total matter profiles converge well down to much
smaller radii than $r_{\rm c}$ implying that this limit is very
conservative. This is a consequence of comparing stacked rather than
individual halos since the stacks tend to average deviations arising
from the additional mass scales represented in the high resolution
simulation. We conclude from this analysis that the total matter and
dark matter profiles of stacked halos are well converged in our
simulations and that we can draw robust conclusions about their
properties for $r>r_{\rm c}$ in both the L100N1504 and L025N0752
simulations.

The gas profiles in these simulations display a much poorer level of
convergence. The disagreement between the two simulations increases at
radii larger than $r>r_{\rm c}$. However, since the mass in gas is
negligible at all radii and at all halo masses, the poor convergence
of the gas profiles does not affect our conclusions regarding the dark
and total matter profiles. We defer the question of the convergence of
gaseous profiles to future studies and simulations.

%-------------------------------------------------------------------------------

\subsection{Stacked halo density and cumulative mass of relaxed halos}
\label{ssec:density_profiles}

Having established a robust convergence criterion for stacked halos we
now analyse their profiles extracting halos of mass
$M_{200}\geq10^{11}\msun$ from the L100N1504 simulation and halos of
mass $10^{10}\msun \leq M_{200} \leq 10^{11}\msun$ from the
L025N0376 simulation.

\begin{figure*}
\includegraphics[width=\textwidth]{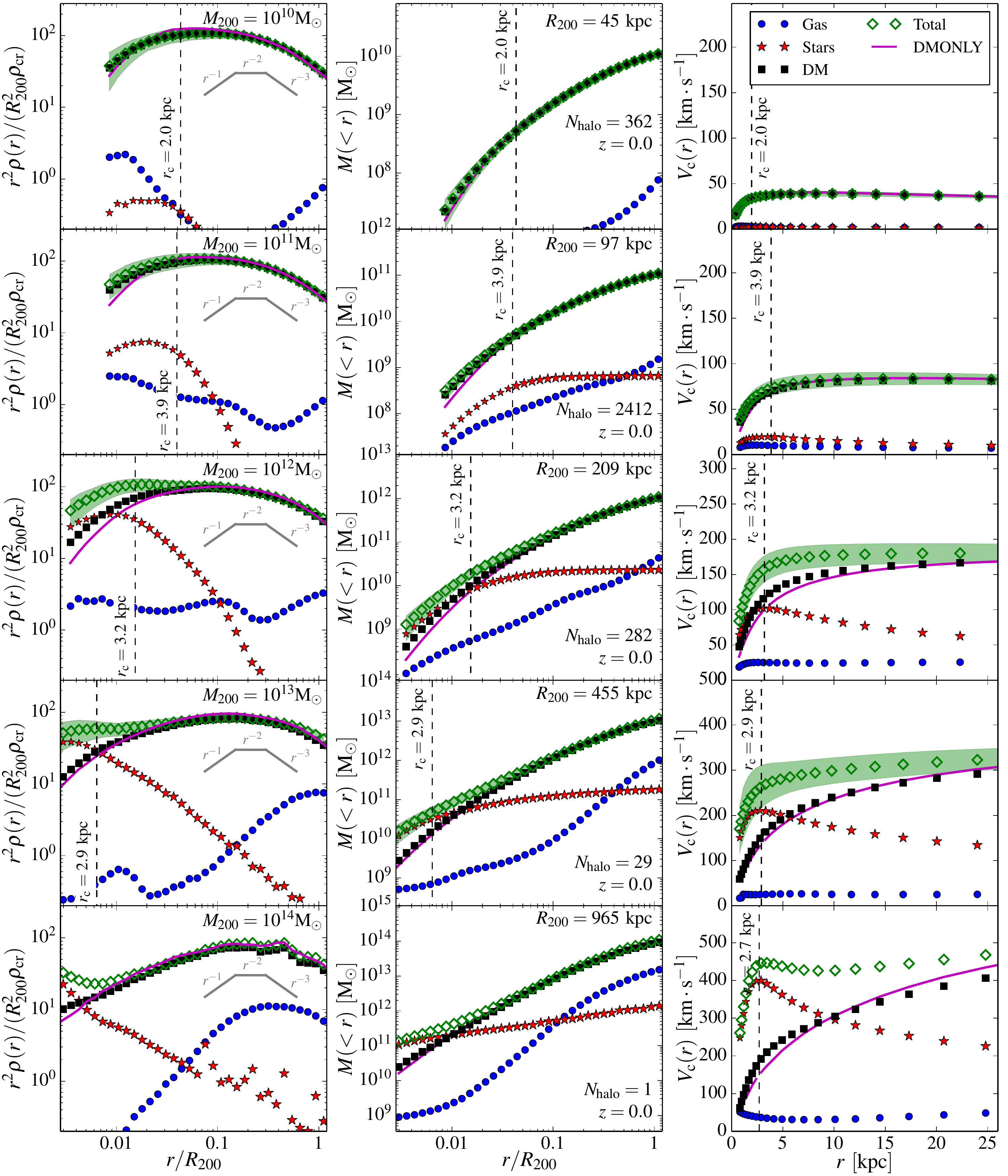}
\caption{From left to right: the density, mass and circular velocity
  profiles for stacks of relaxed halos in different mass bins at
  $z=0$.  From top to bottom: bins centred on
  $M_{200}\approx10^{10}\msun$, $10^{11}\msun$, $10^{12}\msun$,
  $10^{13}\msun$ and $10^{14}\msun$. Profiles of the total matter
  (green diamonds), dark matter (black squares), gas (blue circles)
  and stellar component (red stars) are shown for the halos extracted
  from the \eagle simulation.  Profiles extracted from halos of
  similar mass in the \dmonly simulation are shown with a magenta
  solid line on all panels. The \rms scatter of the total profile is
  shown as a green shaded region. The vertical dashed line shows the
  (conservative) resolution limit, $r_{\rm{c}}$, introduced in the
  previous subsection; data are only shown at radii larger than the
  force softening. The number of halos in each mass bin is indicated
  in the middle panel of each row. The density profiles have been
  multiplied by $r^2$ and normalized to reduce the dynamic range of
  the plot and to enable easier comparisons between different halo
  masses. Note that following the analysis of
  Section~\ref{ssec:HaloMass}, matched halos are not guaranteed to
  fall into the same mass bin.  The oscillations seen in the profiles
  of the two highest mass bins, which have only a few examples, are
  due to the object-to-object scatter and the presence of
  substructures.}
\label{fig:profilesComponent}
\end{figure*}

Fig.~\ref{fig:profilesComponent} shows the stacked profiles for five
different halo mass bins.  The left-hand column shows that the DM is
the dominant component of the density of halos of all masses outside
about one percent of $R_{200}$.  Inside this radius the stellar
component begins to contribute and even dominate in the case of halos
with mass $\gtrsim10^{12}\msun$.  Considering only the baryonic
matter, the inner radii are dominated by stars, but gas dominates
outside of $\sim0.1R_{200}$, as we already saw in
Fig.~\ref{fig:stellarFraction}.  In halos of Milky Way size
($M_{200}\sim10^{12}\msun$) the density profile of the gas is roughly
isothermal with $\rho(r)\propto r^{-2}$.  The stars exhibit a steep
profile, $\rho(r)\propto r^{-3} - r^{-4}$, in the region where this is
resolved ($r>r_{\rm{c}}$). The resolution of our simulations is not
sufficient to enable the discussion of the stellar profile in the
central part of the galaxies, within $\sim3~\rm{kpc}$ of the centre of
potential.

The shape of the dark matter profiles in the \eagle simulation are
typically very close to those obtained in the \dmonly simulation.  The
profiles depart from the \dmonly shape in halos with
$M_{200}\gtrsim10^{12}\msun$, where the slope in the inner regions
(below $0.1R_{200}$) is slightly steeper.  This indicates that some
contraction of the dark matter has taken place, presumably induced by
the presence of baryons in the central region.

The {\em total} density profiles of the \eagle halos also closely
resemble those of the \dmonly simulation. This follows because the DM
dominates over the baryons at almost all radii. In halos with a
significant stellar fraction, the total profile is dominated by the
stars within $\sim0.01R_{200}$. This creates a total inner profile
that is steeper than in the \dmonly simulations. The stellar
contribution is dominant only in the first few kiloparsecs almost
independently of the halo mass.  Given that \dmonly halos have
profiles similar to an NFW profile, this implies that the total
profile will be closer to an NFW for more massive halos because the
stars will only be important inside a smaller fraction of the virial
radius.  This is most clearly seen in the $10^{14}\msun$ halo where
the profile is dominated by the DM and follows the NFW form down to
$~0.01R_{200}$. Similarly, in the smallest halos,
$M_{200}\approx10^{10}\msun$, the baryon content is so low that the
total matter profile behaves almost exactly like the dark matter
profile and is hence in very good agreement with dark matter-only
simulations.

It is also interesting to note the absence in our simulations of DM
cores of size $0.5-2~\rm{kpc}$ such as have been claimed in
simulations of individual halos of various masses, assuming different
subgrid models and, in some cases, different techniques for solving
the hydrodynamical equations \citep[e.g.][]{Navarro1996b, Read2005,
  Mashchenko2006, PontzenGovernato2012, Teyssier2013, Martizzi2013,
  Arraki2014,
  PontzenGovernato2014,Trujillo2015,Murante2015,Onorbe2015}, even
though such cores would have been resolved in our highest resolution
simulations. As first shown by \cite{Navarro1996b}, density cores can
be generated by explosive events in the central regions of halos when
gas has become self-gravitating. Our simulations include violent
feedback processes but these are not strong enough to generate a core
or even a systematic flattening of the inner DM profile on resolved
scales. We cannot, of course, rule out the possibility that the
central profile could be modified even with our assumed subgrid model
in higher resolution simulations.

%-------------------------------------------------------------------------------

\subsection{Halo circular velocities}
\label{ssec:rotation_curves}

The right-hand column of Fig.~\ref{fig:profilesComponent} shows the
rotation curves. Those for Milky Way mass halos display a flat profile
at radii greater than $10~\rm{kpc}$ as observed in our galaxy and
others \citep[e.g.][]{Reyes2011}. The dominant contribution of the DM
is clearly seen here. The stellar component affects only the first few
kiloparsecs of the rotation curve. The rotation curves of halos with a
significant ($>0.01$) stellar fraction (i.e.  halos with
$M_{200}>3\times10^{11}\msun$) have a higher amplitude than the
corresponding \dmonly stacked curves at small radii
$r\lesssim10~\rm{kpc}$. The combination of the stellar component and
contraction of the inner dark matter halo leads to a maximum rotation
speed that is $\approx30\%$ higher in the \eagle simulation compared
to that in \dmonly.

To assess whether the circular velocity profiles for the galaxies in the
\eagle simulation are realistic, we compare them to a sample of
observed disc galaxies. We use the data from \cite{Reyes2011}, who
observed a sample of 189 spiral galaxies and used $\rm{H}\alpha$ lines
to measure the circular speeds. From their SDSS $r-$band
magnitudes and $g-r$ colours, we derive the stellar masses of their
galaxies using the $M_*/L$ scaling relation of  \cite{Bell2003}. We
apply a $-0.1~\rm{dex}$ correction to adjust these stellar mass
estimates from their assumed `diet Salpeter' IMF to our adopted 
\cite{Chabrier2003} IMF, and apply the correction from
\cite{Dutton2011} to convert our masses to the MPA/JHU definitions 
(See \cite{McCarthy2012} for the details.).

\begin{figure*}
\includegraphics[width=0.9\textwidth]{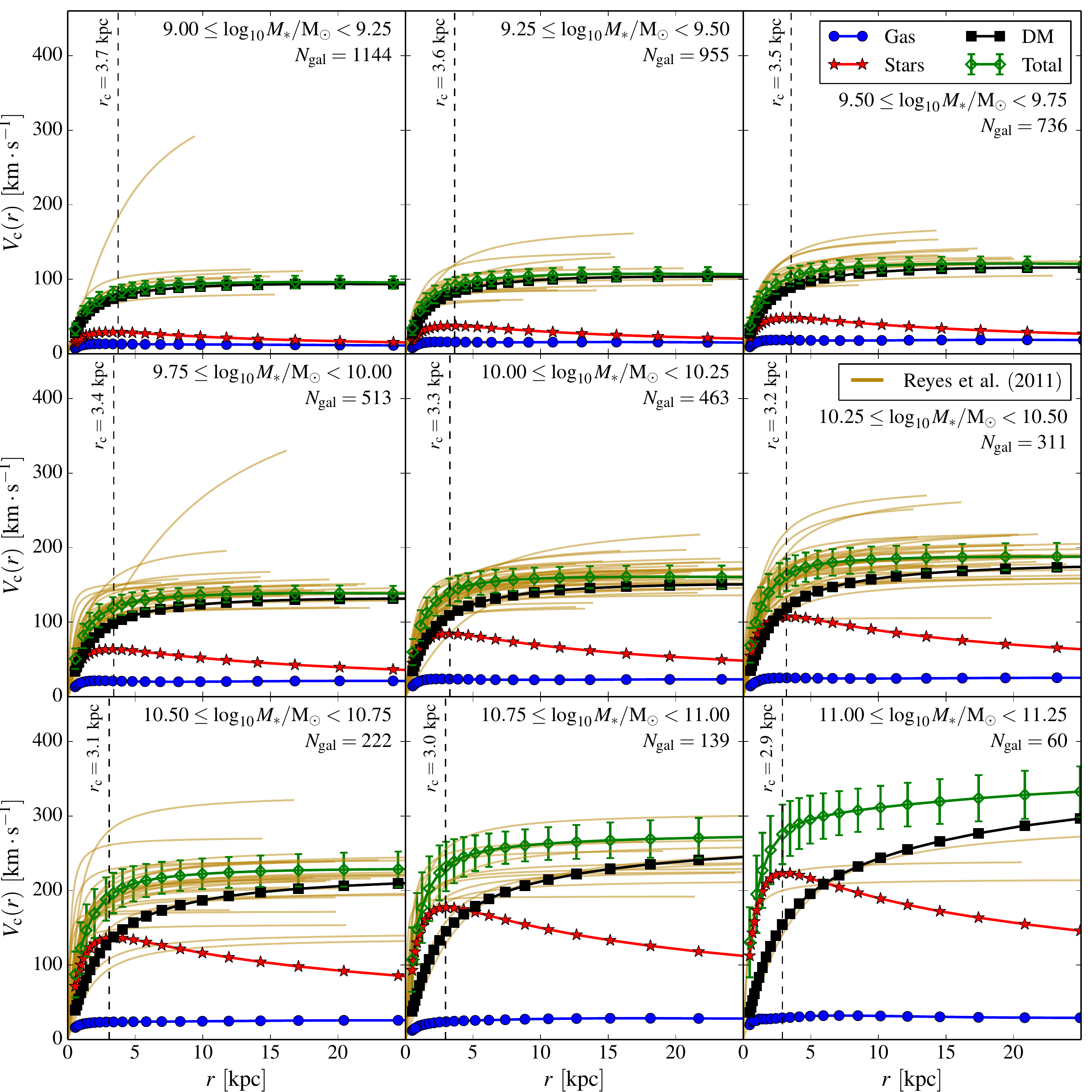}
\caption{Simulated circular velocity curves and observed spiral galaxy
rotation curves in different stellar mass bins. The green diamonds
with error bars correspond to the total circular velocity and
the \rms scatter around the mean. The black squares,
red stars and blue circles represent the mean contributions of dark
matter, star and gas particles respectively. The dashed vertical line
is the conservative resolution limit, $r_{\rm{c}}$. The background brown
curves are the best-fit $\rm{H}\alpha$ rotation curves extracted from
\citet{Reyes2011}.  We plot their data
up to their $i-$band measured isophotal $R_{80}$ radii.
} 
\label{fig:rotationCurves}
\end{figure*}

In Fig.~\ref{fig:rotationCurves} we show the rotation curves of our
sample of relaxed halos binned by the stellar mass contained within an
aperture of $30~\rm{kpc}$, as used by \cite{Schaye2014} who already
compared the predicted maximum circular velocities to
observations. The simulated galaxies match the observations
exceptionally well, both in terms of the shape and the normalisation
of the curves.  For all mass bins up to $M_*<10^{11}\msun$, the \eagle
galaxies lie well within the scatter in the data. Both the shape and
the amplitude of the rotation curves are reproduced in the
simulation. The scatter appears to be larger in the real than in the
simulated population, particularly in the range $10.5 < \log_{10}
M_*/\msun < 10.75$ (lower left panel), but the outliers in the data
might affected by systematic errors \citep{Reyes2011} arising, for
instance, from the exact position of the slit used to measure spectral
features or from orientation uncertainties.

The rotation curves for the highest stellar mass bin in the
simulation, $M_* >10^{11}\msun$, show a clear discrepancy with the
data.  Although the general shape of the curves is still consistent,
the normalisation is too high.  Part of this discrepancy might be due
to the selection of objects entering into this mass bin. The data
refer to spiral galaxies, whereas no selection besides stellar mass
has been applied to the sample of simulated halos. This highest mass
bin is dominated by elliptical objects in \eagle. Selecting
spiral-like objects (in a larger simulation) may well change the
results at these high stellar masses. A more careful measurement of
the rotation velocities in the simulations in a way that is closer to
observational estimates (e.g. by performing mock observations of
stellar emission lines) might also reduce the discrepancies. We defer
this, more careful, comparison to future work.

At all masses beyond the convergence radius the dominant contribution
to the rotation curve comes from the dark matter. For the highest mass
bins the stellar contribution is very important near the centre and
this is crucial in making the galaxy rotation curves relatively flat.
As already seen in the previous figure, the contribution of gas is
negligible.

%-------------------------------------------------------------------------------

\subsection{An empirical universal density profile}
\label{ssec:profiles_fit}

It is well known that the density profiles of relaxed halos extracted
from dark matter only simulations are well fit by the NFW profile
(Eqn.~\ref{eq:nfw}) at all redshifts down to a few percent of the
virial radius \citep{Navarro1997,Bullock2001, Eke2001, Navarro2004,
  Shaw2006, Maccio2007, Neto2007, Duffy2008, Ludlow2013, Dutton2014}.
The total matter profiles shown in Fig.~\ref{fig:profilesComponent}
for the \eagle simulation follow the NFW prediction in the outer
parts, but the inner profile is significantly steeper than the NFW
form, which has an inner slope ($\rho(r\rightarrow0) = r^{-\eta}$ with
$\eta\approx1$).  The deviations from an NFW profile can be quite
large on small scales.

\begin{figure*}
\includegraphics
[width=\textwidth]{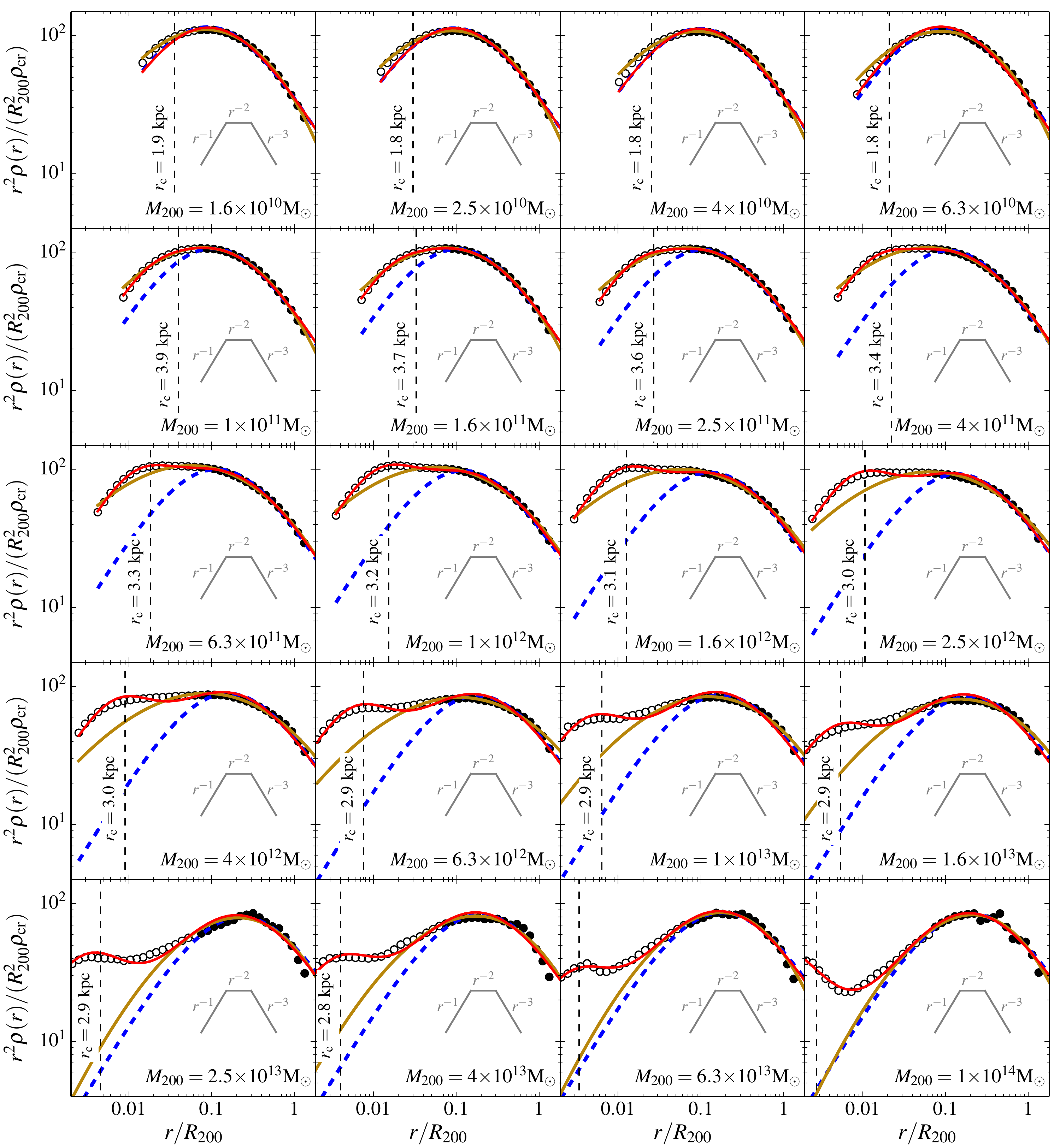}
\caption{Stacked density profiles of the total mass normalized by the
  average $R_{200}$ radius and scaled by $r^2$ for halos of different
  masses. The filled circles are the data points used to fit an NFW
  profile following \citet{Neto2007}, i.e. radial bins above
 data points below it are shown using fainter symbols. The blue
  dashed lines correspond to the NFW fit to the filled circles, while
  the brown lines correspond to an Einasto profile fit to all radial
  bins down to the convergence radius, $r_{\rm{c}}$.  The red solid
  line is the best-fit profile given by Eqn.~\ref{eq:densityProfile},
  which includes an NFW contribution for the outer parts of the halos
  and an additional contribution around the centre to model the
  baryons. The best-fitting parameters for each mass bins are given in
Table~\ref{tab:bestFitParameters}.}
\label{fig:profiles}
\end{figure*}

To show this, we fit the total mass profiles using the fitting
procedure defined by \cite{Neto2007}.  We fit an NFW profile to the
stacked profiles over the radial range $[0.05,1]R_{\rm vir}$, shown
respectively as blue dashed curves and filled circles in
Fig.~\ref{fig:profiles}. This choice of minimum radius is larger than
the conservative convergence radius given by version of the \cite{Power2003}
criterion that we adopted in the previous section. As described in
Section~\ref{ssec:density_profiles}, the bins are spherical and spaced
logarithmically in radius.

The \cite{Neto2007} fit is performed by minimizing a $\chi^2$
expression with two free parameters, $r_{\rm{s}}$ and
$\delta_{\rm{c}}$, characterising the NFW profile, over a set of $N_{\rm b}
(=17)$ radial bins. We use the Levenberg \& Marquart method to
minimize the \rms deviation, $\sigma_{\rm{fit}}$, between the binned
logarithmic densities $\rho_{\rm{i}}$ and the NFW profile
$\rho_{\rm{NFW}}$:
\begin{equation}
 \sigma_{\rm{fit}} = \frac{1}{N_{\rm{b}}-1} \sum_{i=1}^{N_{\rm{b}}} 
\left(\log_{10}\rho_{\rm{i}} - 
\log_{10}\rho_{\rm{NFW}}(\delta_{\rm{c}},r_{\rm{s}})\right)^2.
\label{eq:chi2}
\end{equation}
Note that the bins are weighted equally.

The best-fit profile for each stacked halo mass bin is shown in
Fig.~\ref{fig:profiles} as a blue dashed line. The NFW profile is a
very good fit to the filled circles, confirming that the outer parts
of the halos are well described by this profile within $R_{200}$.
However, the NFW profile is clearly a poor fit at small radii
($r\lesssim0.05R_{\rm vir}$) for halos with a significant stellar
mass, i.e. for halos above $\sim3\times10^{11}\msun$, as expected from
Fig.~\ref{fig:profilesComponent}, due to the increased contribution of
the stars and the subsequent contraction of the DM profile. For halo
masses above $10^{12}\msun$, the discrepancy between the NFW
prediction and the actual total mass density profile reaches factors
of two close to the resolution limit .

When multiplied by $r^2$, the NFW profile reaches a maximum at
$r=r_{\rm{s}}$. For $M_{200}>3\times10^{11}\msun$ the profiles do not
display a single sharp maximum but rather a broad range of radii at
almost constant $r^2\rho(r)$, i.e. a quasi isothermal profile. For
$M_{200} \gtrsim3\times10^{13}\msun$, the difference is even more
striking as a second maximum appears at small radii. We will explore
alternative fitting formula in what follow, but it is clear that a
fitting formula describing the most massive halos will require several
parameters to work well.

In their detailed study, \cite{Navarro2004} explored the use of a more
general class of profiles, where the slope varies with radius as a
power law.  This alternative profile was originally introduced by
\cite{Einasto1965} to model old stellar populations in the Milky Way,
and so \cite{Navarro2004} called it the ``Einsasto profile'':

\begin{equation}
 \rho(r) = \rho_{-2} 
\exp\left[-\frac{2}{\alpha}\left(\left(\frac{r}{r_{-2}}\right)^\alpha - 
1\right)\right],
\label{eq:Einasto}
\end{equation}
which can be rewritten as

\begin{equation}
\frac{\rm{d}\ln \rho(r)}{\rm{d}\ln r} = -2\left(\frac{r}{r_{-2}} \right)^\alpha,
\end{equation}
to highlight that the slope is a power-law of radius.
\cite{Navarro2004} showed that halos in \dmonly simulations are
typically better fit by the Einasto profile and that the value of the
power law parameter, $\alpha\approx0.17$, can be used across the whole simulated 
halo mass
range. This was confirmed by \cite{Gao2008} and \cite{Duffy2008} who found a 
weak
dependence of $\alpha$ on the peak-height parameter. \cite{Gao2008}
demonstrated that the Einasto profile is more robust
to choices of the minimal converged radius, $r_{\rm{c}}$, improving the
quality of the fit.

In the case of our sample of halos, the additional freedom to change
the slope of the power law describing the density profile helps
improve the fit. We use the same procedure as in the NFW case to find
the best-fitting parameters $(r_{-2}, \rho_{-2}, \alpha)$ but instead
of using only the radial bins with $r>0.05R_{\rm vir}$, we use all
bins with $r > r_{\rm{c}}$. The number of bins used is now a function
of the halo mass. The resulting best-fit profiles are displayed in
Fig.~\ref{fig:profiles} as solid yellow lines.  The fits are slightly
better than in the NFW case simply because the rolling power law
allows for a wider peak in $r^2\rho(r)$, but the Einasto profile is
clearly unable to capture the complex behaviour seen in the profiles
of the highest mass bins.  The better fit quality is only incidental.
Furthermore, if we had used the full range of radial bins for the NFW
fitting procedure, we would have obtained similar fits as the two
functions are very similar.  Similarly, restricting the Einasto fit to
the bins with $r>0.05R_{\rm vir}$ yields a best fit profile (and
$\sigma_{\rm{fit}}$) almost identical to the NFW ones shown by the
dashed blue lines.

Clearly, in the presence of baryons, neither the NFW nor the Einasto
profile faithfully represents the inner matter density profile. As
Fig.~\ref{fig:profilesComponent} showed, the inner profile is shaped
by both a substantial stellar contribution and the contraction of the
dark matter associated with the elevated baryon fraction towards the
centre. We find that the total profile can be fit everywhere by the
following formula:
\begin{equation}
 \frac{\rho(r)}{\rhocr} = 
\frac{\delta_{\rm{c}}}{\left({r}/r_{\rm{s}}\right)\left(1 + 
{r}/{r_{\rm{s}}}\right)^2} 
+ 
\frac{\delta_{\rm{i}}}{\left({r}/{r_{\rm{i}}}\right)\left(1+\left({r}/{r_{\rm{i}
}}\right)^2\right)}.
\label{eq:densityProfile}
\end{equation}
The first term is the NFW profile, which we have shown gives a good
fit to the outer, DM-dominated profile. The second term is
NFW-like in that is shares the same asymptotic behaviour at small and
large radii and has a slope of -2 at its scale radius, $r=r_{\rm{i}}$. We
have found by trial and error that its sharper transition relative to
the NFW profile between the asymptotic slope regimes of -1 and -3,
which causes it to rise a factor of two above a corresponding NFW
profile that shares the same scale radius and asymptotic behaviour at
small and large radii, make it particularly suitable for describing the
deviations in the density profiles above an NFW profile seen in the
central regions of the \eagle halos.

We fit this profile using all the radial bins down to our resolution
limit, $r_{\rm{c}}$. We rewrite expression (\ref{eq:chi2}) using our
new profile and minimize $\sigma_{\rm{fit}}$ leaving the four
parameters $(r_{\rm{s}}, \delta_{\rm{c}}, r_{\rm{i}},
\delta_{\rm{i}})$ free.  The resulting fits are displayed in
Fig.~\ref{fig:profiles} as red solid lines. The values of the
best-fitting parameters are given in
Table~\ref{tab:bestFitParameters}. The fit is clearly of a much better
quality than the NFW and Einasto formulas for the same set of radial
bins.

For the lowest mass halos ($M_{200}<6\times10^{10}\msun$), this new profile does 
not provide a better $\sigma_{\rm 
fit}$ than a standard NFW profile does. This is expected since the baryons have 
had little impact on their inner 
structure. The values of $r_{\rm i}$ and $\delta_{\rm i}$ are, hence, not 
constrained by the fits. For these low mass 
stacks, we only provide the best-fitting NFW parameters in 
Table~\ref{tab:bestFitParameters} instead of the parameters 
of our alternative profile.

The different features of the simulated halos are well captured by the
additional component of our profile. We will demonstrate in the next
sections that the additional degrees of freedom can be recast as
physically meaningful quantities and that these are closely correlated
with the halo mass. As in the case of the NFW profile, this implies
that this new profile is effectively a one parameter fit, where the
values of all the four parameters depend solely on the mass of the
halo.  It is worth mentioning that this profile also reproduces the
trends in the radial bins below the resolution limit $r_{\rm{c}}$. \\

\begin{table*}
\begin{minipage}{137mm}
  \caption{Best-fit parameters for the profile
    (Eqn.~\ref{eq:densityProfile}) for each stack of relaxed halos as
    plotted in Fig.~\ref{fig:profiles}. The tabulated values
    correspond to the black circles plotted in
    Figs.~\ref{fig:new_concentration}, \ref{fig:coreSize} and
    \ref{fig:coreMass}. The first column gives the centre of the mass
    bin used for each stack and the last column the number of
    halos in each of the stacks. The concentration, $c_{200}$,
    and inner profile mass, $M_{\rm{i}}$, are defined, respectively, by
    Eqns.~\ref{eq:defconc} and~\ref{eq:M_i}. For the halo stacks in the lowest 
mass bins, the profile 
\ref{eq:densityProfile} does not provide a better fit than a standard NFW. We 
hence only give the best-fitting 
parameters to the NFW fit. }
\label{tab:bestFitParameters}
\begin{tabular}{|r|r|r|r|r|r|r|r|r|r|}

$M_{200}~[\msun]$ & $R_{200}~[\rm{kpc}]$ & $r_s~[\rm{kpc}]$ & $c_{200}~[-]$ 
&$\delta_c~[-]$ & $r_i~[\rm{kpc}]$ & 
$\delta_i~[-]$ & $M_i~[\msun]$ & $N_{\rm{halo}}$\\
\hline
$  1\times10^{10}$ & $ 45.4$ &  $ 4.2$ & $10.7$ & $5.2\times10^{4}$  & $   
\textendash$ & $   \textendash$ & $ 
\textendash$ & $362$\\
$1.6\times10^{10}$ & $ 52.8$ &  $ 4.8$ & $11.0$ & $5.5\times10^{4}$  & $   
\textendash$ & $   \textendash$ & $   
\textendash$ & $231$\\
$2.5\times10^{10}$ & $ 61.4$ &  $ 5.7$ & $10.7$ & $5.2\times10^{4}$  & $   
\textendash$ & $   \textendash$ & $   
\textendash$ & $153$\\
$  4\times10^{10}$ & $ 70.8$ &  $ 6.7$ & $10.5$ & $  5\times10^{4}$  & $   
\textendash$ & $   \textendash$ & $   
\textendash$ & $96$\\
$6.3\times10^{10}$ & $ 83.5$ &  $ 9.8$ & $8.5$ & $2.7\times10^{4}$  & $2.01$ & 
$1.25\times10^{5}$ & $5.66\times10^{8}$ & 
$96$\\
$  1\times10^{11}$ & $ 97.4$ &  $11.7$ & $8.3$ & $2.5\times10^{4}$  & $2.23$ & 
$1.53\times10^{5}$ & $9.44\times10^{8}$ 
& $2412$\\
$1.6\times10^{11}$ & $113.7$ &  $14.1$ & $8.0$ & $2.3\times10^{4}$  & $2.38$ & 
$2.12\times10^{5}$ & $1.58\times10^{9}$ 
& $1657$\\
$2.5\times10^{11}$ & $132.6$ &  $17.2$ & $7.7$ & $2.1\times10^{4}$  & $2.59$ & 
$2.85\times10^{5}$ & $2.74\times10^{9}$ 
& $1119$\\
$  4\times10^{11}$ & $154.3$ &  $20.6$ & $7.5$ & $1.9\times10^{4}$  & $2.56$ & 
$4.75\times10^{5}$ & $4.45\times10^{9}$ 
& $681$\\
$6.3\times10^{11}$ & $180.3$ &  $25.7$ & $7.0$ & $1.6\times10^{4}$  & $2.61$ & 
$7.28\times10^{5}$ & $7.17\times10^{9}$ 
& $457$\\
$  1\times10^{12}$ & $208.8$ &  $31.7$ & $6.6$ & $1.4\times10^{4}$  & $2.78$ & 
$9.22\times10^{5}$ & $ 
1.1\times10^{10}$ & $282$\\
$1.6\times10^{12}$ & $244.7$ &  $38.3$ & $6.4$ & $1.3\times10^{4}$  & $2.89$ & 
$1.18\times10^{6}$ & 
$1.58\times10^{10}$ & $180$\\
$2.5\times10^{12}$ & $286.3$ &  $44.3$ & $6.5$ & $1.4\times10^{4}$  & $2.73$ & 
$1.72\times10^{6}$ & 
$1.94\times10^{10}$ & $126$\\
$  4\times10^{12}$ & $332.4$ &  $54.2$ & $6.1$ & $1.3\times10^{4}$  & $2.65$ & 
$2.17\times10^{6}$ & 
$2.23\times10^{10}$ & $83$\\
$6.3\times10^{12}$ & $386.6$ &  $68.6$ & $5.6$ & $1.1\times10^{4}$  & $2.55$ & 
$2.85\times10^{6}$ & 
$2.63\times10^{10}$ & $60$\\
$  1\times10^{13}$ & $455.2$ &  $73.0$ & $6.2$ & $1.4\times10^{4}$  & $2.26$ & $ 
4.2\times10^{6}$ & $ 
2.7\times10^{10}$ & $29$\\
$1.6\times10^{13}$ & $534.3$ &  $95.3$ & $5.6$ & $1.1\times10^{4}$  & $2.82$ & 
$3.16\times10^{6}$ & 
$3.95\times10^{10}$ & $27$\\
$2.5\times10^{13}$ & $631.4$ &  $130.0$ & $4.9$ & $7.7\times10^{3}$  & $2.13$ & 
$6.81\times10^{6}$ & 
$3.65\times10^{10}$ & $5$\\
$  4\times10^{13}$ & $698.9$ &  $124.6$ & $5.6$ & $1.1\times10^{4}$  & $2.81$ & 
$4.32\times10^{6}$ & 
$5.31\times10^{10}$ & $8$\\
$6.3\times10^{13}$ & $838.1$ &  $141.7$ & $5.9$ & $1.2\times10^{4}$  & $2.73$ & 
$5.23\times10^{6}$ & 
$5.87\times10^{10}$ & $4$\\
$  1\times10^{14}$ & $964.7$ &  $188.1$ & $5.1$ & $8.9\times10^{3}$  & $0.909$ & 
$1.05\times10^{8}$ & 
$4.38\times10^{10}$ & $1$\\

\hline

\end{tabular}
\end{minipage}
\end{table*}

For completeness, we give the analytic expressions for both the
enclosed mass, $M(r<R)$, and the gravitational potential, $\Phi(r)$,
for the empirical profile of Eqn.~\ref{eq:densityProfile},
\begin{eqnarray}
 M(r<R) &=& 2\pi\rhocr\Bigg(2\delta_{\rm{c}} r_{\rm{s}}^3
\left[\ln\left(1+\frac{R}{r_{\rm{s}}}\right)-\frac{R}{R+r_{\rm{s}}}\right]
\nonumber \\
 && + \delta_{\rm{i}} 
r_{\rm{i}}^3\ln\left(1+\frac{R^2}{r_{\rm{i}}^2}\right)\Bigg),
\label{eq:massProfile}
\end{eqnarray}
and
\begin{eqnarray}
 \Phi(r)&=&-4\pi G \rhocr \Bigg(\frac{\delta_{\rm{c}} r_{\rm{s}}^3}{r}
\ln\left({\displaystyle 1+\frac{r}{r_{\rm{s}}}}\right) \\ 
&&+\delta_{\rm{i}}r_{\rm{i}}^2\left[\frac{\pi}{2}-\arctan\left(\frac{r}{r_{\rm{i
}}}\right) + 
\frac{r_{\rm{i}}}{2r}\ln\left(1+\frac{r^2}{r_{\rm{i}}^2}\right)\right]\Bigg). 
\nonumber
 \label{eq:potentialProfile}
\end{eqnarray}
The expressions for an NFW profile are recovered by setting $\delta_{\rm{i}}=0$.

Finally, we stress that while this function provides an excellent fit
to the results over the range of applicability the second term should
not be interpreted as a description of the stellar profile.  Rather,
the second term models a combination of the effect of all components,
including the contraction of the dark matter, and is only valid above
our resolution limit which is well outside the stellar half-mass
radius.  Higher-resolution simulations, with improved subgrid models,
would be needed to model accurately the stars and gas in these very
inner regions.

%-------------------------------------------------------------------------------

\subsection{Dark matter density profile}
\label{ssec:DM_profiles}
 
\begin{figure*}
\includegraphics[width=\textwidth]{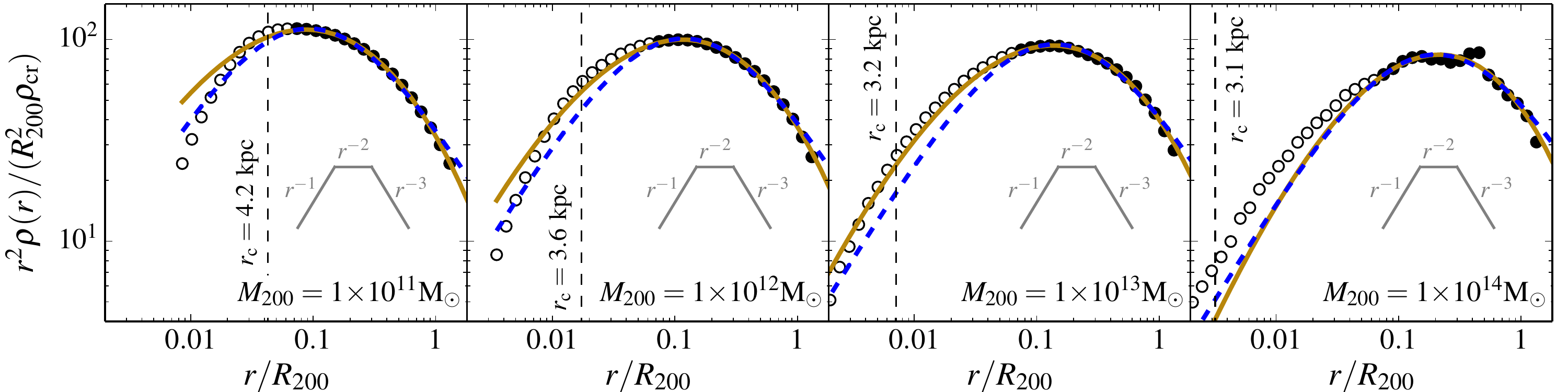}
\caption{Stacked density profiles of the \dmonly halos normalized by
the average $R_{200}$ radius and scaled by $r^2$ for a selection of
masses. The filled circles are the data points used to fit an NFW
profile following \citet{Neto2007}.  The vertical line shows the
resolution limit. Data points are only shown at radii larger than the 
Plummer-equivalent 
softening ($2.8\epsilon=0.7~\rm{kpc}$). The blue dashed and solid brown lines
correspond, respectively, to the 
best-fit NFW and Einasto profiles to the filled circles.  Only one halo 
contributes to the right hand
panel.}
\label{fig:DMONLY_profile}
\end{figure*} 

It is interesting to see whether the radial distribution of dark
matter is different in the \dmonly and \eagle simulations.  In this
subsection we look at the density profiles of just the DM in both the
\dmonly and \eagle simulations.  In Fig.~\ref{fig:DMONLY_profile} we
show the profiles of the stacked halos extracted from the \dmonly
simulation for different halo mass bins. The dark matter outside 
$0.05R_{\rm{vir}}$ is well fit by the NFW profile, in agreement with
previous work. The yellow curves show the best fit Einasto profile,
and in agreement with many authors \citep{Navarro2004, Gao2008,
  Dutton2014} we find that the Einasto fit, with one extra parameter,
provides a significantly better fit to the inner profile.

\begin{figure*}
\includegraphics[width=\textwidth]{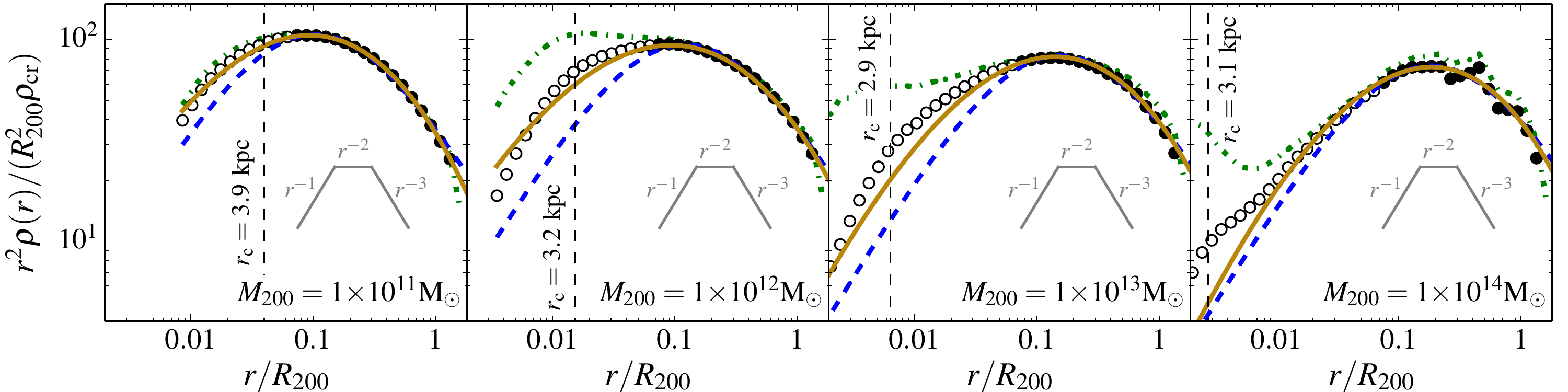}
\caption{Stacked density profiles of the dark matter component of the
  \eagle halos normalized by the average $R_{200}$ radius and scaled
  by $r^2$ for a selection of halo masses. The green dash dotted line
  represents the total mass profile (from Fig.~\ref{fig:profiles}.
  The vertical line shows the resolution limit. Data points are only shown at 
radii larger than the Plummer-equivalent 
softening ($2.8\epsilon=0.7~\rm{kpc}$). The blue dashed
  lines and solid brown lines correspond, respectively, to the
  best-fit NFW and Einasto profiles to the filled circles.}
\label{fig:EAGLE_DM_profile}
\end{figure*}

 We show the stacked DM density profiles for the \eagle
simulation in Fig.~\ref{fig:EAGLE_DM_profile} together with NFW 
and Einasto fits to the density at $0.05 \leq r/R_{\rm{vir}} \leq 1$. For 
the radii beyond $0.05R_{\rm{vir}}$ the NFW profile provides a
good fit.  The Einasto profile fits are better in the
inner regions, but for the middle two mass bins
($10^{12}\msun$ and $10^{13}\msun$), the DM profile
rises significantly above the Einasto fit.  This rise coincides with a
more pronounced feature in the total mass profile. The peak of
the central stellar mass fraction occurs at this same halo mass
scale, as shown in Fig.~\ref{fig:stellarFractionCentre}.

We conclude that the DM components of our simulated halos in both the
\dmonly and \eagle simulations are well described by an NFW profile
for radii $[0.05R_{200}-R_{200}]$.  For the \dmonly simulation an
Einasto profile provides a better fit than an NFW profile at smaller
radii. However, for the \eagle simulation neither an NFW nor the
Einasto profile provide a particularly good fit inside $0.05R_{\rm
  vir}$ for halos in the $10^{12}\msun$ and $10^{13}\msun$ mass bins,
where the contribution of stars to the inner profile is maximum. For
less massive and more massive halos than this both functions give
acceptable fits.

In their detailed study of ten simulated galaxies from the MaGICC
project \citep{Stinson2013}, \cite{DiCintio2014} fitted
$(\alpha,\beta,\gamma)$-profiles \citep{Jaffe1983} to the DM profiles
of haloes in the mass range $10^{10}\msun \leq M_{\rm
  vir}\leq10^{12}\msun$ and studied the dependence of the parameters
on the stellar fraction. We leave the study of the DM profiles in the
\eagle halos to future work but we note that although in the small
halo regime, $M_{200}\leq10^{12}\msun$, an
$(\alpha,\beta,\gamma)$-profile may be a good fit, the profiles of our
most massive halos, $M_{200}\geq10^{13}\msun$, show varying slopes
down to small radii, $r\leq0.05R_{\rm vir}$, and are unlikely to be
well fit by such a function as was already suggested by \cite{DiCintio2014}.
 
%-------------------------------------------------------------------------------

\subsection{Halo concentrations}
\label{ssec:concentrations}

The concentration of a halo, $c_{\rm{X}}$, is conventionally defined
by the ratio, $c_{\rm{X}} =R_{\rm{X}}/r_{\rm conc}$, where
$R_{\rm{X}}$ is the radius within which mean internal density is
$X\rhocr$, and $r_{\rm conc}$ is the radius at which the spherically
averaged density profile (assumed monotonic) obeys
\begin{equation}
\frac{{\rm d}\ln\rho(r)}{{\rm d}\ln r} = -2. 
\label{eq:defconc}
\end{equation}
For an NFW profile, $r_{\rm conc}=r_{\rm{s}}$, while for an Einasto
profile $r_{\rm conc}=r_{-2}$.  We set $\rm{X}=200$.

  Previous work \citep{Navarro1997, AvilaReese1999, Jing2000,
Bullock2001, Eke2001, Zhao2003,Neto2007, Maccio2007, Duffy2008, Gao2008,
Dutton2014} has shown that the concentration and the mass of relaxed
halos are anticorrelated (at $z=0$), and follow a power law of the
form
\begin{equation}
c_{200} = A\left(\frac{M_{200}}{ 10^{14} h^{-1} \msun}\right)^B,
\label{eq:massConcentration}
\end{equation}
where $A\approx5$ and $B\approx-0.1$. The best-fit values of these
parameters are sensitive to the cosmological parameters, particularly
to the values of $\sigma_8$ and $\Omega_{\rm{m}}$
\citep[e.g.][]{Duffy2008,Dutton2014}.  The value of $c_{200}$ at
redshift zero is linked to the background density of the Universe at
the time of formation of the halo \citep{Navarro1997, Ludlow2013}
which is affected by $\sigma_8$ and $\Omega_{\rm{m}}$. Higher values
of these parameters lead to earlier halo formation times at a given
mass and therefore higher concentrations.  The concentrations of
individual halos of a given mass scatter about the median value with
an approximately log-normal distribution \citep{Jing2000, Neto2007}.
The amplitude of this scatter decreases with halo mass \citep{Neto2007}.

\begin{figure}
\includegraphics
[width=\columnwidth]{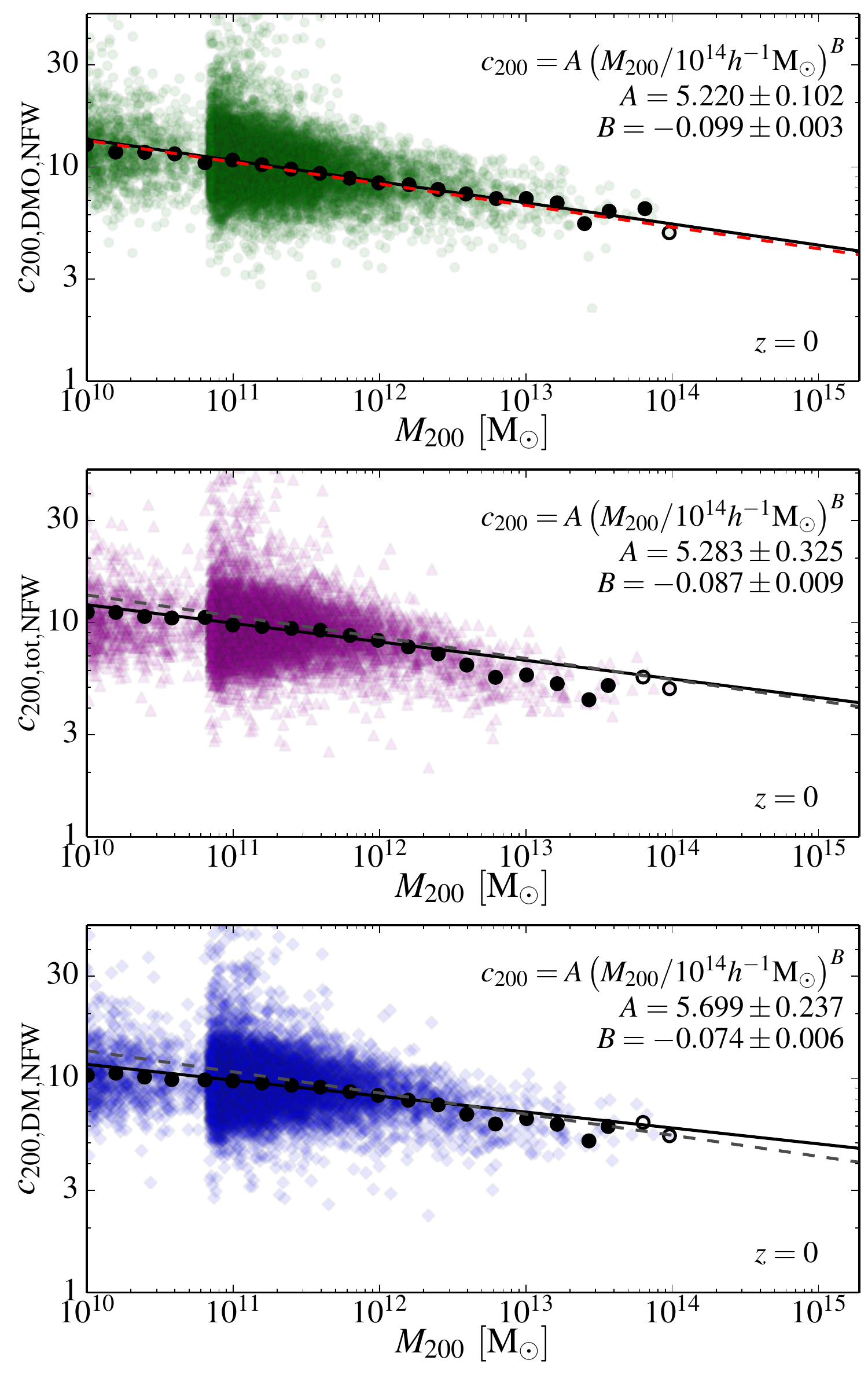}
\caption{Halo concentration, $c_{200}$, as a function of mass
  $M_{200}$.  The top panel shows the \dmonly simulation fit with the
  canonical NFW profile over the range $[0.05-1]R_{\rm vir}$.  The
  middle panel shows the same fit applied to the total matter density
  profiles of the \eagle halos.  The bottom panel shows the same fit
  to just the dark matter in the \eagle halos. The faint coloured
  points in each panel are the values for individual halos and the
  black circles the values for the stacked profiles in each mass
  bin. Halos and stacks with $M_{200}<6\times10^{10}\msun$ are taken from the 
  L025N0752 simulation whilst the higher mass objects have been extracted from 
the L100N1504 simulation.
  The solid black line is the best-fit power law
  (Eqn.~\ref{eq:massConcentration}) to the solid black circles. The
  best-fit parameters are shown in each panel. The best-fit power law
  to the \dmonly halos is repeated in the other panels as a dashed
  line. The red dashed line on the first panel is the best-fit
  relation from \citet{Dutton2014}.}
\label{fig:concentration}
\end{figure}

While formally Eqn.~\ref{eq:defconc} implicitly defines $R_{\rm conc}$, it is
impractical to apply a differential measure of the density to
determine the concentrations of individual halos, even in simulations,
because the density profiles are noisy and sensitive to the presence
of substructures. In practice, the concentration is determined by
fitting the spherically averaged density profile over a range of radii
encompassing $r_{\rm{s}}$ with a model. This approach only works if
the model provides a good description of the true halo profile over
the fitted range.  We have shown in Section~\ref{ssec:profiles_fit}
that the density profiles of halos in both the \eagle and DMO
simulations are well described by an NFW profile over the range
$[0.05-1]R_{\rm vir}$, so we fit an NFW model over this range.

Fig.~\ref{fig:concentration} shows the NFW concentration of relaxed
halos as a function of halo mass for the \dmonly and \eagle
simulations. The top panel shows the \dmonly simulation. The black
line is the best fit power law of Eqn.~\ref{eq:massConcentration} to
the solid black circles (corresponding to the stacks containing at
least five halos) using Poissonian errors for each bin. We have
verified that fitting individual halos (faint green circles in the
same figure) returns essentially the same values of $A$ and
$B$. Table~\ref{tab:massConcentration} lists the best-fitting values
of these parameters. It is worth mentioning that the best-fitting
power laws fit the halo stacks in the simulations equally well.

\begin{table}
  \caption{Best fitting parameters and their $1\sigma$ uncertainty for the
    mass-concentration relation (Eqn.~\ref{eq:massConcentration}) of
    the stacks of relaxed halos. The values correspond to those shown in
    the legends in Fig.~\ref{fig:concentration}. From top to bottom: NFW fit to the
    \dmonly halos, NFW fit to the total mass of the \eagle halos, and NFW
    fit to the dark matter component of the \eagle halos. All profiles were fit over
    the radial range $[0.05-1]R_{\rm vir}$. The uncertainties are
    taken to be the diagonal elements of the correlation matrix 
    of the least-squares fitting procedure.}
\label{tab:massConcentration}
\begin{center}
\begin{tabular}{|l|c|c|}
 Fit & $A$ & $B$\\
\hline
$c_{200, \rm{DMO}}$ & $ 5.22\pm0.10 $ & $ -0.099\pm0.003 $ \\
$c_{200, \rm{tot}, \rm{NFW}}$ & $5.283\pm0.33$ & $-0.087\pm0.009 $ \\
$c_{200, \rm{DM}, \rm{NFW}}$ & $5.699\pm0.24 $ & $-0.074\pm0.006$ \\
\end{tabular}
\end{center}
\end{table}

The mass-concentration relation of \cite{Dutton2014} is shown as a red
dashed line in the top panel of Fig.~\ref{fig:concentration}.  This
fit is based on a series of \dmonly cosmological simulations of a
\lcdm model very similar to ours with the cosmological parameters
values taken from the \cite{Planck2013} data.  Using several
volumes at different resolutions, they were able to determine the
concentration-mass relation over the range $10^{10}\msun < M_{200} <
1.5\cdot10^{15}\msun$ at $z=0$. Fitting an NFW model to estimate the
concentration, as we do here, they obtained
\begin{equation}
c_{200} = 5.05 \left(\frac{M_{200}}{ 10^{14}h^{-1}\msun}\right)^{-0.101},\nonumber
\end{equation} 
which agrees well with our results.

Not unexpectedly, given the sensitivity of the concentration to
changes in the cosmological parameters, the values for the fit we
obtain for the \dmonly simulation are significantly different from
those reported by \cite{Neto2007}, \cite{Maccio2007} and
\cite{Duffy2008}. Compared to the latter, the slope ($B$) is
steeper and the normalisation ($A$) is higher. This change can be
attributed mainly to changes in the adopted cosmological
parameters $(\sigma_8,\Omega_{\rm{m}})$ which were $(0.796,0,258)$ in
\cite{Duffy2008} and $(0.8288,0.307)$ here.

\begin{figure}
\includegraphics[width=\columnwidth]{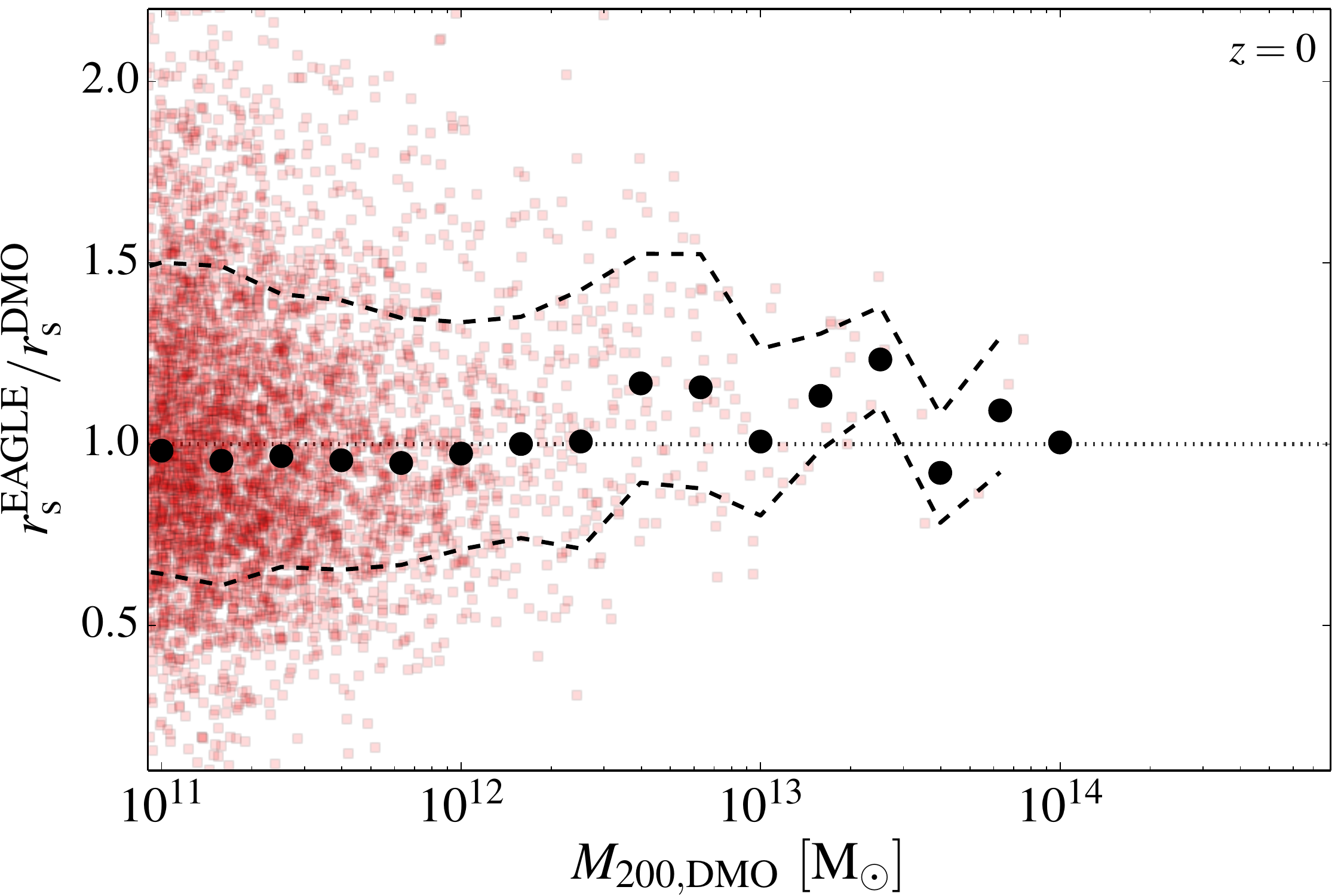}
\caption{Ratio of NFW scale radii, $r_{\rm{s}}$, in matched relaxed
  halos in the \dmonly and \eagle simulations. The black points are
  placed at the geometric mean of the ratios in each mass bin.}
\label{fig:scale_ratios}
\end{figure}

The second panel of Fig.~\ref{fig:concentration} shows the
concentrations for the total matter density profiles of the \eagle
simulation obtained using the same fitting procedure. The best-fitting
parameters for the mass - concentration relation are given in the
second line of Table~\ref{tab:massConcentration}.  Both the amplitude
and slope are consistent with the values for the \dmonly
simulation. As discussed in Section \ref{ssec:HaloMass}, matched halos
in the \dmonly and \eagle simulations have, on average, a lower mass
in the \eagle simulation. For the smallest halos, the average ratio is as low
as $0.72$. Because of this shift in mass, some difference in the
concentration-mass relation might be expected between the two
simulations but, since the value of the slope is small and
$0.72^{-0.1} \simeq 1.04$, the effect on the amplitude is also
small. A consequence of the shift in $M_{200}$ is that the relative
sizes of $R_{200}$ for matched halos is
$R_{200}^{\rm{EAGLE}}/R_{200}^{\rm{DMO}} \simeq0.9$. In
Fig.~\ref{fig:scale_ratios} we show that the mean ratio of
$r_{\rm{s}}^{\rm{EAGLE}}/r_{\rm{s}}^{\rm{DMO}}$ for matched relaxed
halos is also slightly below unity, so the net effect of those two
shifts is that the concentrations are very similar in both
simulations.

Finally, the bottom panel of Fig.~\ref{fig:concentration} shows the
concentration of the DM only component of \eagle halos. We
fit an NFW profile in the same way as for the total matter profiles in
the panels above.  As would be expected from the analysis of
Fig.~\ref{fig:profiles} and the fact that the outer parts of the
dark halos are well described by the NFW profile, the same trend
with mass can be seen as for the \dmonly simulation. The best-fitting
power law to the mass-concentration relation is given at the bottom of
Table~\ref{tab:massConcentration}. The values of the parameters are
again close to the ones obtained for both the \eagle and the \dmonly
simulations.

We stress that the agreement between the \eagle and \dmonly simulations breaks
down if we include radii smaller than $0.05R_{\rm{vir}}$ in the fit. Hence, the 
mass - concentration relation given for \eagle in Table \ref{tab:massConcentration} 
should only be used to infer the density profiles beyond $0.05R_{\rm{vir}}$.

%------------------------------------------------------------------------------------------------
\subsection{Best-fit parameter values for the new density profile}
\label{ssec:fit_parameters}

\begin{figure}
\includegraphics[width=\columnwidth]{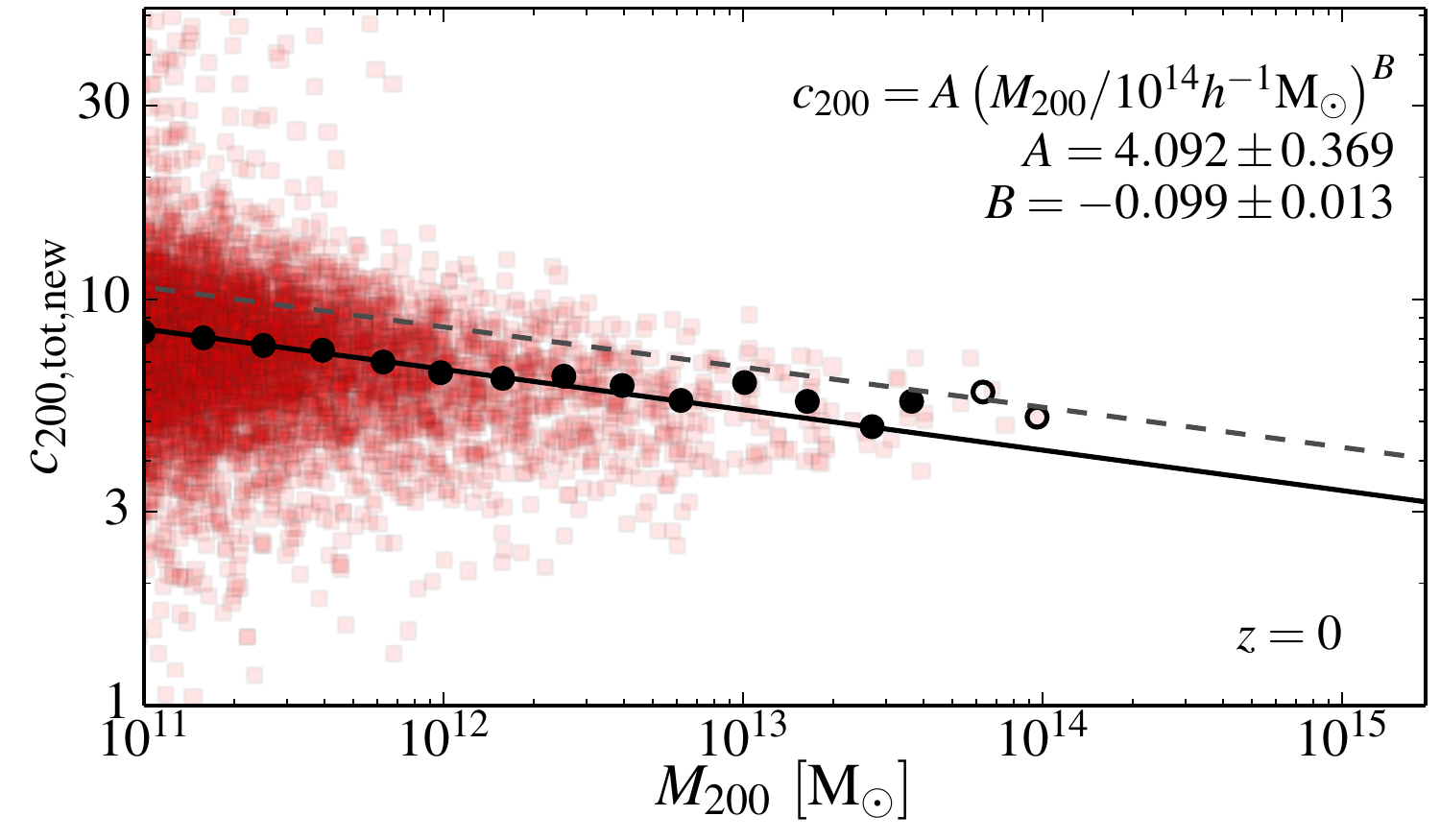}
\caption{Halo concentration, $c_{200}$, as a function of mass,
  $M_{200}$, for the total matter density profiles of the \eagle
  simulation using the fitting function of
  Eqn.~\ref{eq:densityProfile} and the $r_{\rm{s}}$ parameter to
  define the concentration, $c_{200} = R_{200}/r_{\rm{s}}$.  The
  colour points are for individual halos and the black circles for the
  stacked profiles in each mass bin. The solid black line is the
  best-fit power law (Eqn.~\ref{eq:massConcentration}) to the solid
  black circles. The best-fit values are given in the legend at the
  top right. The dashed line shows the best fitting power law to the
  halos extracted from the \dmonly simulation fitted using an NFW
  profile.}
\label{fig:new_concentration}
\end{figure}

We showed in Section~\ref{ssec:profiles_fit} that the density profiles
of halos in the \eagle simulation are not well fit by an NFW profile
in the inner regions, and we proposed Eqn.~\ref{eq:densityProfile} as
a new fitting formula for these profiles. This new profile has two
lengthscales, $r_{\rm{s}}$ and $r_{\rm{i}}$, where the former
describes the NFW-like outer parts of the halo, and the latter the
deviations from NFW in the inner regions.  For lower-mass halos these
two lengths become similar, so both terms of the profile can contribute
significantly to the density at all radii.  We can still define the
concentration of a halo in this model as $R_{200}/r_{\rm{s}}$, but we
would expect to obtain a different mass-concentration relation from
that for the dark matter-only case.  Fig.~\ref{fig:new_concentration}
shows this relation for relaxed \eagle halos. The anticorrelation seen
when fitting an NFW profile is still present and we can use the same
power-law formulation to describe the mass-concentration relation of
our halo stacks.  The values of the best-fit parameters, given in the
figure, differ significantly from those obtained using the NFW fits
listed in Table~\ref{tab:massConcentration}.

We now consider the two remaining parameters of the profile described
by Eqn.~\ref{eq:densityProfile}. The inner component is characterized
by two quantities, a scale radius, $r_{\rm{i}}$, and a density
contrast, $\delta_{\rm{i}}$. We stress that this inner profile
should not be interpreted as the true underlying model of the galaxy
at the centre of the halo. It is an empirical model that describes the
deviation from NFW due to the presence of stars and some contraction
of the dark matter. The profiles have been fit using the procedure
described in Section~\ref{ssec:profiles_fit} using all radial bins
with $r>r_{\rm{c}}$.

\begin{figure}
\includegraphics[width=\columnwidth]{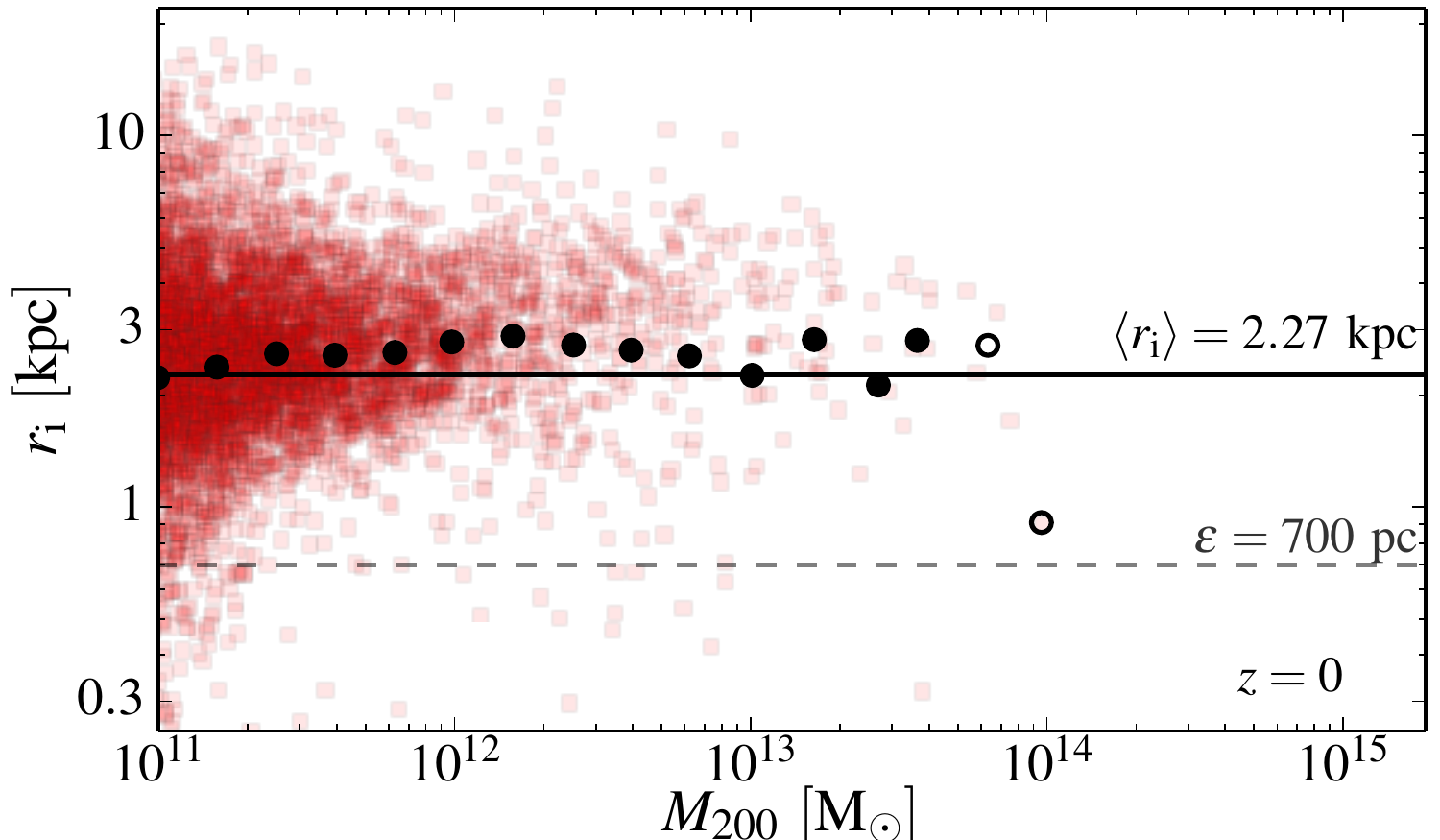}
\caption{The characteristic radius, $r_{\rm{i}}$, of the central
  component as function of halo mass (Eqn.~\ref{eq:densityProfile})
  for halos in the \eagle simulation.  The red squares correspond to
  all the halos fitted individually and the overlaying black circles
  to the stacked halos in each mass bin. Stacks containing less than
  three objects are shown as open circles. The minimum Plummer-equivalent softening
  length ($\epsilon=0.7~\rm{kpc}$) is indicated by the grey dashed
  line at the bottom of the figure.  The average value of the stacks
  with more than three objects is indicated by a solid black line.}
\label{fig:coreSize}
\end{figure}

The dependence of the $r_{\rm{i}}$ scale radius on the halo mass is
shown in Fig.~\ref{fig:coreSize}.  The radius $r_{\rm{i}}$ is roughly
constant over the entire halo mass range in the simulation. The
scatter is large at all masses, but there is a weak trend with mass in
the low-mass regime. This regime is, however, difficult to study as
may be seen in the first few panels of Fig.~\ref{fig:profiles}: for
the smallest halos, the effects due to baryons are small and the
profile is thus closer to NFW than for the higher-mass bins.

The empirical profile (Eqn.~\ref{eq:densityProfile}) tends towards an
NFW profile as $\delta_{\rm{i}}\rightarrow0$ or
$r_{\rm{i}}\rightarrow0$.  We find that, for the smallest halos, there
is a degeneracy between these two parameters and the values of
$r_{\rm{i}}$ and $\delta_{\rm{i}}$ can be changed by an order of
magnitude (self-consistently) without yielding a significantly
different $\sigma_{\rm{fit}}$ value.  This is not a failure of the
method but rather a sign that the baryonic effects on the profile
shape become negligible for the lowest-mass halo, at least for the
range of radii resolved in this study.

Rather than working with the $\delta_{\rm{i}}$ and $r_{\rm{i}}$
parameters, we can combine them into a single parameter that reflects
the additional mass contained in the central parts of the halo above
and above that from the NFW component. Integrating the inner
profile up to $r_{\rm{i}}$, we can obtain an estimate of this
additional mass which we define as:
\begin{equation}
 M_{\rm{i}} = (2\pi\ln2)\rhocr r_{\rm{i}}^3\delta_{\rm{i}} \approx 4.355\rhocr r_{\rm{i}}^3\delta_{\rm{i}} 
\label{eq:M_i}.
\end{equation}
If $r_{\rm{i}}$ were really constant, then $M_{\rm{i}}$ would simply
be a proxy for $\delta_{\rm{i}}$.

\begin{figure}
\includegraphics[width=\columnwidth]{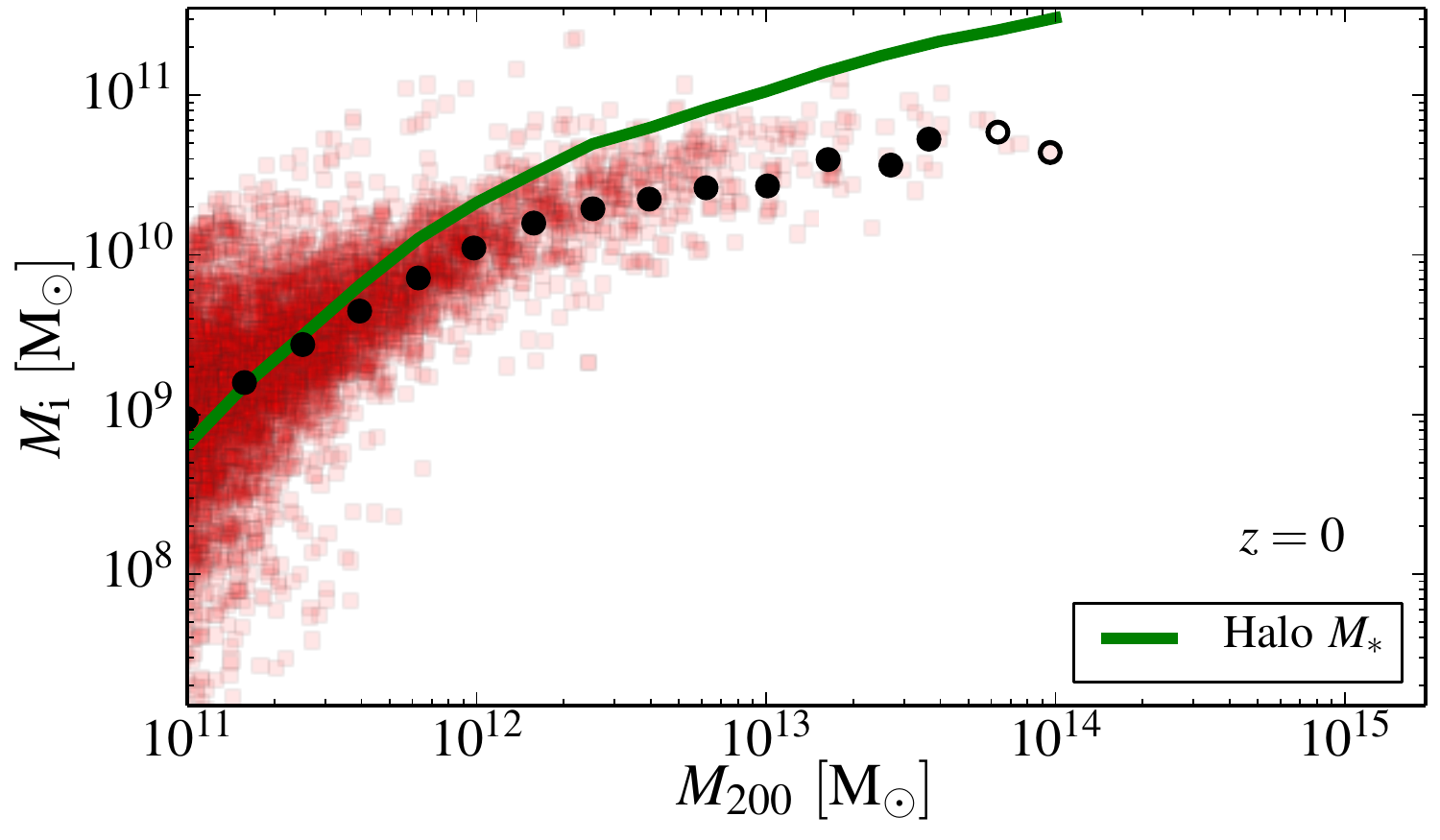}
\caption{The mass, $M_{\rm{i}}$, defined in Eqn.~\ref{eq:M_i}, as a
  function of halo mass, $M_{200}$. The red squares correspond to the
  individual halos and the overlaying black circles to the stacked
  profiles.  The green solid line is the stellar mass - halo mass
  relation from the \eagle simulation \citep{Schaye2014}.}
\label{fig:coreMass}
\end{figure}

The mass, $M_{\rm{i}}$, is shown in Fig.~\ref{fig:coreMass} as a
function of the halo mass, $M_{200}$.  The black points corresponding
to the stacked profiles lie in the middle of the relation for
individual halos.  The mass, $M_{\rm{i}}$, increases with halo
mass. For halos with $M_{200}\lesssim10^{12}\msun$, the fraction,
$M_{\rm{i}}/M_{200}$, increases with $M_{200}$ highlighting that the
effect of the baryons is more important for the bigger halos. This
could have been expected by a careful inspection of
Fig.~\ref{fig:stellarFractionCentre}, which shows that the central
stellar and baryonic fractions peak at
$M_{200}\approx10^{12}\msun$. For larger halos, the
$M_{200}$-$M_{\rm{i}}$ relation flattens reflecting the decrease in
stellar fractions seen at the centre of the largest \eagle halos.

To confirm this conjecture, we plot the stellar mass - halo mass
relation for the \eagle simulation as a solid green line in the same
figure \citep{Schaye2014}\footnote{Note that the \eagle simulation
  reproduces abundance matching results \citep{Schaye2014}.}.
Neglecting the two highest mass bins (open circles), the similarity
between this relation and our somewhat arbitrary definition of
$M_{\rm{i}}$ seems to indicate that the stellar mass of the halos is
related to this parameter.  The definition of the mass, $M_{\rm{i}}$,
could also be modified to take into account other properties of the
galaxy in the halo. We could, for instance, include the galaxy size
(half-stellar mass radius or half-light radius, for example) instead
of $r_{\rm{i}}$ in the definition of $M_{\rm{i}}$. It would then be
interesting to see how this newly defined mass correlates with the
galaxy's stellar mass.

\subsection{A non-parametric estimate of the concentration}
\label{ssec:enc_ratio}

The definition of concentration assumes that the halos are well fit by
an NFW (or other) profile. This is the case for our sample of halos
down to radii $\sim0.05R_{\rm vir}$, so we can safely compute the
concentration of these halos as $r_{\rm{s}} > 0.05 R_{\rm vir}$ for
almost all cases of interest. It is nevertheless worthwhile measuring
a proxy for the concentration which does not rely on a specific
parametrization of the profile. This is also more convenient for
observational purposes, where a large range of radii are not always
available to perform a fit. A simpler estimator of the concentration
can then help constrain the models without making assumptions about
the exact shape of the profile.  This is particularly true for X-ray
observations because it is difficult to detect X-ray emission all the
way to the virial radius.

\begin{figure}
\includegraphics[width=\columnwidth]{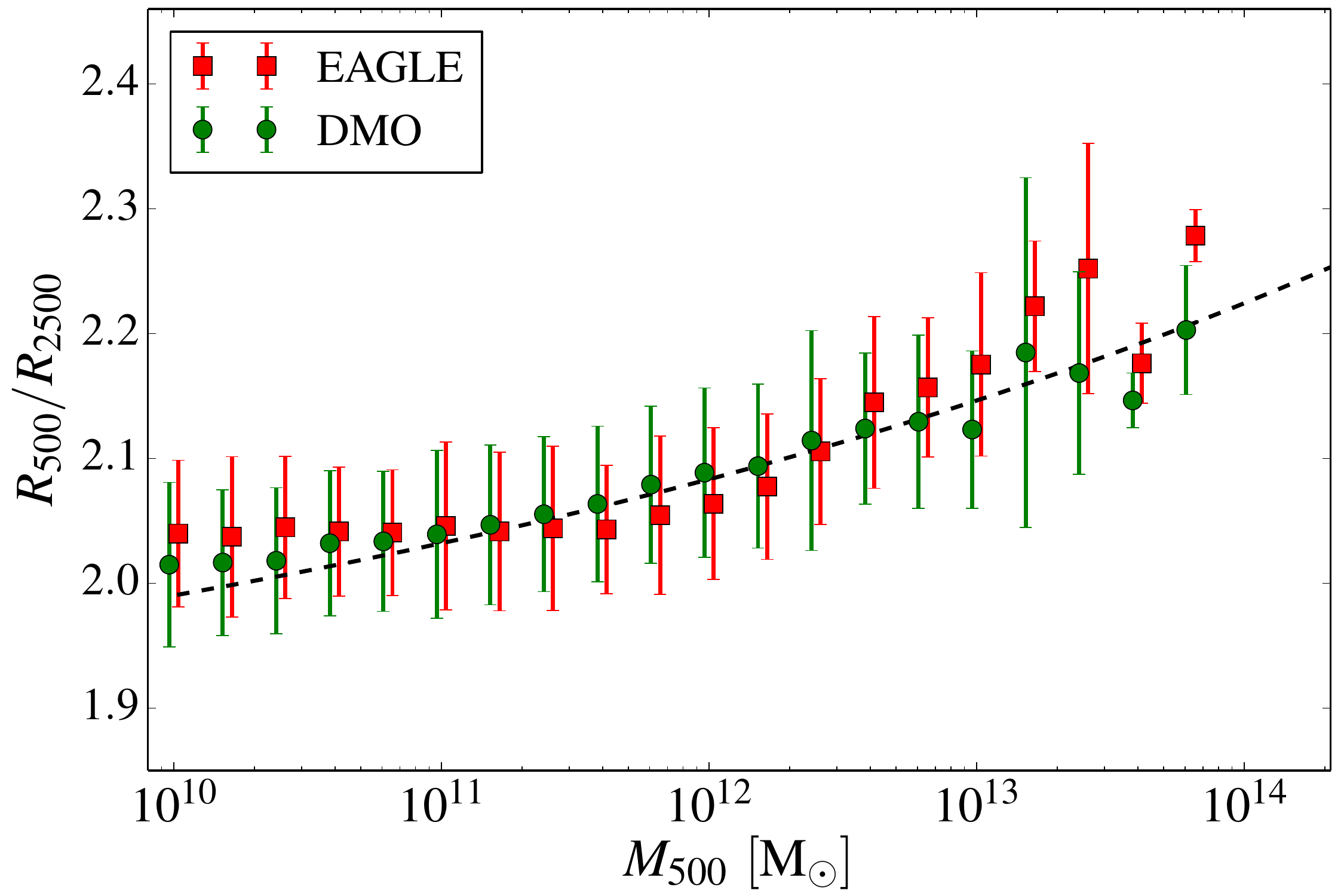}
\caption{The average ratio of the $R_{500}$ and $R_{2500}$ radii as a
  function of halo mass, $M_{500}$, for both the \eagle (red squares)
  and \dmonly (green circles) simulations.The error bars represent the
  $1\sigma$ scatter in the population. To ease the reading of the
  plot, the points with error bars have been artificially displaced by
  $0.02\rm{dex}$ towards the left and right for the \eagle and \dmonly
  results respectively. The black dashed line shows the expected
  relation for a NFW profile with the concentration-mass relation
  determined for the \eagle simulation in
  Section~\ref{ssec:concentrations}.}
\label{fig:nonParametric}
\end{figure}

Such an estimator is given by the ratio of spherical over-density
radii $R_{500}/R_{2500}$ \citep[e.g.][]{Duffy2010}. Both of these
quantities can be obtained without assuming anything about the slope
and functional form of the matter density profile. We show the value
of this ratio as a function of the spherical enclosed mass, $M_{500}$,
in Fig.~\ref{fig:nonParametric}.  The \eagle and \dmonly simulations
show the same trends and the differences between them are smaller than
the scatter between individual halos.  As could already be seen from
the profiles in Figs.~\ref{fig:profilesComponent} and
\ref{fig:profiles}, the effect of modelling the baryons is
concentrated at small radii, well within $R_{2500}$.

\subsection{Limitations of the subgrid model}
\label{ssec:missing_physics}

The convergence test in subsection \ref{ssec:resolution_test}
demonstrated that the simulation results of interest here are
converged at radii, $r>r_{\rm c}$ (given by a modified version of the
criterion proposed by \citealt{Power2003}) and that even at smaller
radii the profiles of stacked halos remain remarkably similar when the
resolution is increased by a factor of $8$ in particle mass. A halo of
mass $M_{200}\approx10^{11}\msun$ is then resolved with
$\mathcal{O}(10^5)$ particles and its stellar disk with
$\mathcal{O}(10^3)$ particles, which is enough to sample star
formation histories with good accuracy and obtain a realistic galaxy
population \citep{Schaye2014, Furlong2014, Crain2014}.

An interesting aspect of our simulations is that no halos (of any
mass) develop density cores in their central regions within the
resolved radial range. By contrast, simulations of dwarf and even
larger galaxies by a number of authors produce such cores \citep[see
references in Sec.~\ref{ssec:density_profiles} and] [for a
review]{PontzenGovernato2014}.  As shown by \cite{Navarro1996b}, a
physical mechanism that can produce a flattening of the inner dark
matter density profile is the sudden removal, in a starburst, of gas
that had previously contracted enough to become self-gravitating,
dominate the central gravitational potential and slowly drag dark
matter in.  The subsequent loss of binding energy from the central
regions by the removal of that gas on a timescale shorter than the
local dynamical time causes the dark matter to flow outwards resulting
in a flattening of the profile to a degree that depends on the size
and mass of the self-gravitating gas component.  A variant of this
process is apparently responsible for the formation of cores in
simulations of dwarf galaxy \citep[e.g.][]{Governato2010} and in 
simulations of galaxy clusters (where the source of energy is an AGN;  
\citealt{Martizzi2013}).

An important aspect of the simulations by \cite{Governato2010} is that
the assumed subgrid model adopts a higher density threshold for star
formation ($10-100~m_{\rm H}\cdot\rm{cm}^{-3}$) than we have assumed
in \eagle (a metallicity-dependent threshold with a value of
$0.031~m_{\rm H}\cdot\rm{cm}^{-3}$ at solar metallicity that traces
the density above which a cold, molecular gas phase is expected to be
present; see \cite{Schaye2004,Schaye2014})\footnote{A significant
  number of stars in \eagle, however, form from gas at much higher
  densities that the threshold; see \cite{Crain2014}}. Although even
the high value assumed by \cite{Governato2010} is many orders of
magnitude below the gas density in the star-forming cores of molecular
clouds, it probably allows a substantial amount of gas to dominate the
potential in the central regions prior to a starburst, as required for
the \cite{Navarro1996b} mechanism to operate\footnote{It is unclear
  whether cold, dense star-forming clouds in a multiphase interstellar
  medium \cite{MckeeOstriker1977} would contain enough mass to
  dominate the central potential of the halo.}. 

It is not obvious {\em a priori} which, if any, of the subgrid models
for star formation used to date is more appropriate, but an important
virtue of the \eagle subgrid model is that it leads to a population of
galaxies with properties that agree well with a large set of
observations, from the regime of dwarf galaxies to the regime of
galaxy clusters \citep{Schaye2014, Furlong2014, Crain2014, Sawala2014,
  Schaller2014b}. None of the simulations that produce cores in the
dark matter has yet been able to demonstrate such success. Indeed,
other large cosmological simulations with different subgrid models to
\eagle such as ``Illustris'' do not appear to have produced density
cores either \citep{Vogelsberger2014}.  In any event, the evidence for
the existence of cores in real galaxies is still a matter of lively
debate, with some authors reporting cores \citep[e.g.][]{Salucci2000,
  Swaters2003, Simon2005, Gentile2007, deBlok2008, Kuzio2008,
  Oh2011a}, others reporting cusps \citep[even for some of the same
galaxies, e.g.][]{Adams2014}, and others arguing that current
kinematical data cannot distinguish cores from cusps (at least in the
case of satellites of the Milky Way for which kinematical studies on
resolved stellar populations are possible; \citealt{Strigari2010,
  Strigari2014}).

Finally, we stress that the conclusions in this paper refer only to
radii larger than $r>r_{\rm c} \approx 1.8~\rm{kpc}$. Higher
resolution simulations would be required to test whether our subgrid
model can generate density cores on smaller scales than those resolved
in the present study.

\section{Conclusions}
\label{sec:conclusion}

The aim of this study was to characterize the mass density profiles of
dark matter halos in a cosmological \lcdm simulation, which includes
dark matter and baryons, and in which the resulting galaxy population
has realistic stellar masses and sizes; we also quantified the
differences with halos in a dark matter-only simulation.  We used the
state-of-the-art \eagle simulation from which we selected halos above
$10^{9}\msun$ to study changes in the mass, and above $10^{11}\msun$ 
to study changes in the internal structure. Our results can be
summarized as follows:

\begin{enumerate}
\item The mass, $M_{200}$, of halos is reduced by the inclusion of
  baryons and associated energy feedback effects.  At the low mass
  end, feedback from star formation expels gas and this reduces the total mass,
  radius and growth factor of the halo; the reduction in mass can be as
  large as $30\%$ for halos with $M_{200}\lesssim10^{11}\msun$. This
  reduction is progressively smaller for larger halos as the source of
  feedback shifts from star formation to AGN. In the \eagle simulation
  there is virtually no effect for masses $M_{200}
  \gtrsim10^{14}\msun$, but the exact value of the mass at which this
  happens could be larger if, as suggested by \cite{Schaye2014}, more
  effective AGN feedback is necessary than is present in \eagle. The reduction in
  mass can be described by the double-sigmoid function of
  Eqn.~\ref{eq:meFit}, and the scatter around the mean by the formula
  of Eqn.~\ref{eq:Myfitscatter}.

\item The circular velocity curves of the \eagle halos are in
  excellent agreement with observational data for galaxies with
   stellar mass ranging from $10^9\msun$ to $5\times10^{11}\msun$
  (corresponding to halo masses in the range $10^{11}\lesssim
  M_{200}/\msun\lesssim 10^{13}$).

\item The radial density profiles of \eagle halos over the radial range 
  $[0.05R_{\rm{vir}},R_{\rm{vir}}]$ are very similar to the profiles of
  halos in dark matter-only simulations and are well described by the 
  NFW formula. Halo concentrations estimated by fitting NFW profiles
  over this range are also very similar to the dark matter-only case. 

\item The central regions of halos more massive than $M_{200}
  \gtrsim10^{12}\msun$, on the other hand, are dominated by the
  stellar component. The presence of these baryons causes a
  contraction of the halo, enhancing the density of dark matter in
  this region. The variation in profile shape is greatest for halos in
  the mass range $M_{200}=10^{12}\msun - 10^{13}\msun$ where the
  stellar mass fraction peaks (as does the total baryonic mass
  fraction within $0.05R_{\rm{vir}}$

\item The radial density profiles of the \eagle halos can therefore be
  well fit (over the radial range resolved in the simulation) by the
  sum of an NFW profile, which describes the outer, dark
  matter-dominated regions, and an NFW-like profile with a sharper
  bend, which describes the combined effect of the presence of stars
  and the contraction of the dark matter halo
  (Eqn.~\ref{eq:densityProfile}). Of the two additional parameters
  required in this fit, one, $r_{\rm{i}}$, is approximately constant
  with halo mass, while the other one, the characteristic inner mass
  scale, $M_{\rm{i}}$, scales with halo mass in a similar way to the
  stellar mass of the central galaxy.

\end{enumerate}

The way in which galaxy formation affects the host halos is a problem
that can only be reliably addressed with simulations of the kind we
have described here. However, it is clear that the nature of these
effects is sensitive to the way in which the baryon physics are
implemented, particularly to the subgrid models for feedback from star
formation and AGN. The \eagle simulations have the great advantage
that the subgrid models have been calibrated so that the simulation
reproduces the local galactic stellar mass function as well as the
distribution of galaxy sizes, and they also reproduce a wide variety
of other observations.  This lends a certain degree of credibility to
our results and it would be interesting to compare them with other
simulations that assume different subgrid models but achieve similarly
good matches to observables over a large range of halo masses. A
limited comparison of this kind is carried out in Appendix~A1.

The simulations investigated here do not have enough resolution to
study dwarf galaxies for which there is much discussion regarding the
formation of central cores in the dark matter density distribution
\citep[for a review see][]{PontzenGovernato2014}. However, the related
high resolution simulations of the Local Group by \cite{Sawala2014},
which use essentially the same subgrid models as \eagle, do resolve
dwarfs. The behaviour of these smaller halos simply continues to
smaller masses the trends seen here: the halos become increasingly
dark matter-dominated and remain well described by the NFW profile.

\section*{Acknowledgements}
We are grateful to Lydia Heck and Peter Draper without whose technical expertise and
support this work would have not been possible. RAC is a Royal Society University Research Fellow.
This work was supported in part by an STFC Consolidated grant to Durham University
and by the European Research Council through ERC Grants Cosmiway (GA
267291), GasAroundGalaxies (GA 278594) and Cosmocomp (GA 238356) and
also the Inter-university Attraction Poles Programme initiated by the
Belgian Science Policy Office ([AP P7/08 CHARM]). This work was also
sponsored by the Dutch National Computing Facilities Foundation (NCF),
with financial support from the Netherlands Organization for
Scientific Research (NWO).  The \eagle simulations used the DiRAC Data
Centric system at Durham University, operated by the Institute for
Computational Cosmology on behalf of the STFC DiRAC HPC Facility
(www.dirac.ac.uk). This equipment was funded by BIS National
E-infrastructure capital grant ST/K00042X/1, STFC capital grant
ST/H008519/1, and STFC DiRAC Operations grant ST/K003267/1 and Durham
University. DiRAC is part of the National E-Infrastructure.  We
acknowledge PRACE for resources on the Curie supercomputer in France.

\bibliographystyle{mn2e}
\bibliography{./bibliography}

\appendix

\section{Uncertainties due to the subgrid models}

As discussed by \cite{Schaye2014}, cosmological hydrodynamical
simulations require subgrid models whose parameters have to be
calibrated against a set of observables. In the case of the \eagle
suite of simulations, the observations used are the $z=0$ galaxy
stellar mass function, the galaxy mass-size relation and the stellar 
mass-black hole mass relation. Using only a subset of these observables, 
it is possible to find different values of the subgrid model parameters 
that match the galaxy stellar mass function \citep{Crain2014}. Hence, 
it is important to assess whether the results presented here depend on these
parameters or on the resolution of the simulation.

\subsection{Changes in the AGN model parameters}
\label{ssec:AGN_changes}

One of the models that matched the selected set of observables is the
\eagle model AGNdT9-L050N0752, which is very similar to the \eagle-Ref
model used in the rest of this paper but whose parameters have been
calibrated to match the group gas fractions and X-ray luminosities
better \citep{Schaye2014}. In this model, the galaxy masses and sizes
are very similar to the Ref model and we have verified that the dark
matter halo profiles extracted from that model are very close to the
ones shown in Section~\ref{ssec:density_profiles} for the halo mass
range represented in this simulation ($M\lesssim2\times10^{13}\msun$).

\begin{figure}
\includegraphics[width=\columnwidth]{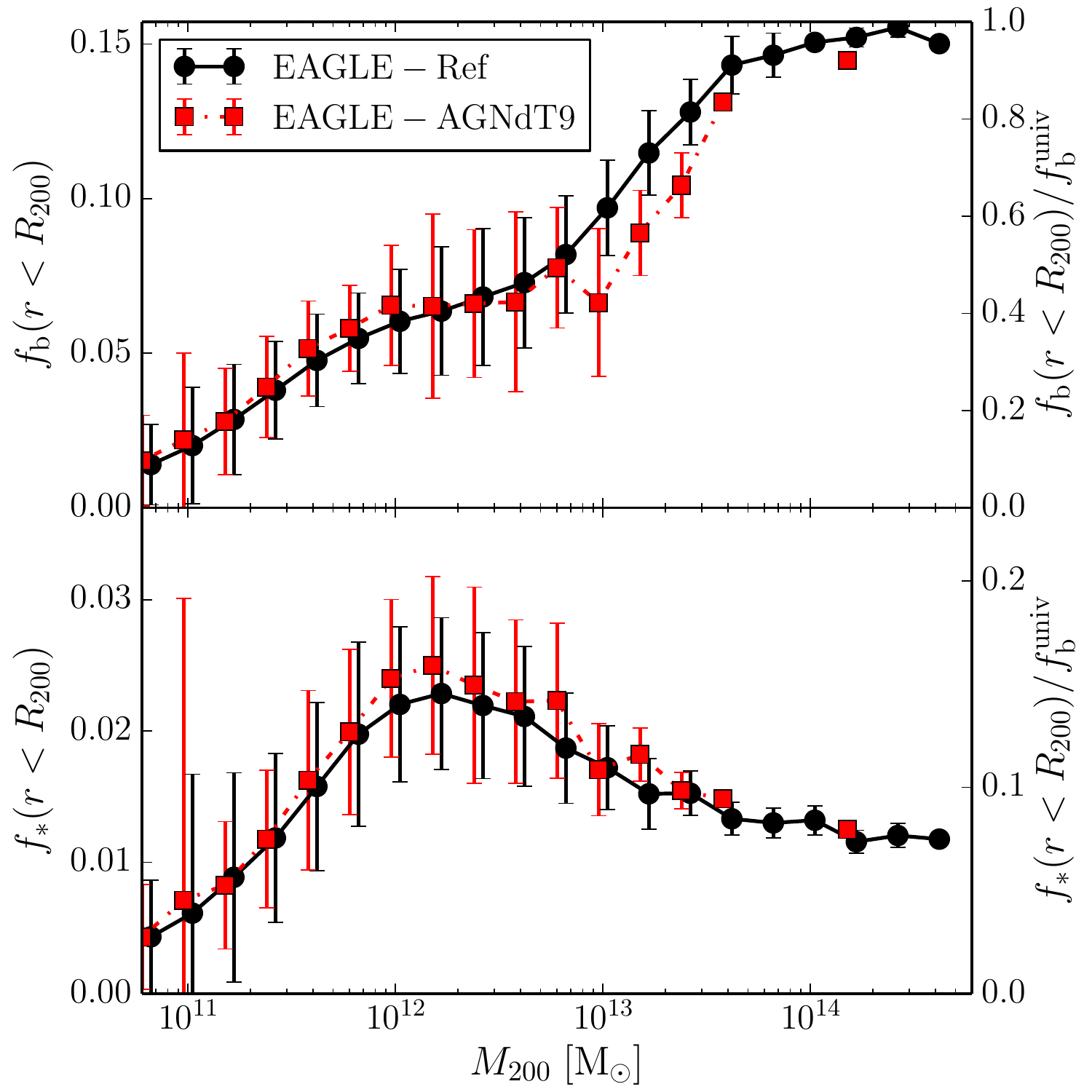}
\caption{Baryon fraction, $f_{\rm b}=M_{\rm b}/M_{200}$ (top panel),
  and stellar fraction, $f_*=M_*/M_{200}$ (bottom panel), within
  $R_{200}$, as a function of halo mass for the \eagle-Ref model
  (black circles) and the \eagle-\mbox{AGNdT9} model (red
  squares). The error bars show the \rms halo-to-halo scatter in each
  mass bin.  The baryon fractions in the halos more massive than
  $10^{13}\msun$ are lower in the \mbox{AGNdT9} model.}
\label{fig:baryonFractionComparison}
\end{figure}

In Section~\ref{ssec:HaloMass} we discussed the difference in halo
masses between the AGNdT9 simulation and its \dmonly equivalent and
showed that the ratio reached unity only for more massive halos than
in \eagle-Ref model. This is, in part, caused by the lower baryon
fractions that these halos
have. Fig.~\ref{fig:baryonFractionComparison} shows the baryon (top
panel) and stellar (bottom panel) fractions for halos extracted from
the \eagle-Ref simulation (black circles) and from the \eagle-AGNdT9
model (red squares). The stellar fractions are comparable in both
models, with any differences laying well within the large halo-to-halo
scatter. The baryon fractions in group-like halos ($10^{13} \msun <
M_{200} < 10^{14}\msun$), however, are systematically lower, by as
much as $20\%$, in the \eagle-AGNdT9 model.  This difference is
reflected in the observed shift in the best fitting parameter,
$M_{23}$, in Eqn.~\ref{eq:meFit} between the two models. The
difference vanishes for the central regions of the halos. The baryon
and stellar fractions within $0.05R_{200}$ are similar in both
simulations indicating that the difference in the AGN treatment has
mostly lead to a change in the structure of the gas outside 
galaxies, impacting on the inferred X-ray luminosities
\citep{Schaye2014}.

\label{lastpage}

\end{document}